# Disorder-free many-body localization in real quantum materials

Yingpeng Qi[1,2#*], Jianmin Yang[3,4#], Zhihui Zhou[2], Qing Xu[2], Yang Lv[2], Xiao Zou[2], Tao Jiang[2], Pengfei Zhu[2], Dongxue Chen[1,5], Zhenrong Sun[6], Lin Xie[7*], Dao Xiang[2,8*], Jiaqing He[3,9*]

[1]Suzhou Institute for Advanced Research, University of Science and Technology of China, Suzhou 215123, China

[2]Key Laboratory for Laser Plasmas (Ministry of Education), School of Physics and Astronomy, Shanghai Jiao Tong University, Shanghai 200240, China.

[3]Department of Physics, State Key Laboratory of Quantum Functional Materials, and Guangdong Basic Research Center of Excellence for Quantum Science, Southern University of Science and Technology, Shenzhen 518055, China.

[4]Department of Materials Science and Engineering, National University of Singapore, 117575, Singapore.

[5]Department of Physics, University of Science and Technology of China, Hefei, Anhui 230026, China

[6]State Key Laboratory of Precision Spectroscopy, and School of Physics and Electronic Science, East China Normal University, Shanghai 200241, China

[7]School of Physical Sciences, Great Bay University, Dongguan 523000, China.

[8]Zhangjiang Institute for Advanced Study, Shanghai Jiao Tong University, Shanghai 201210, China.

[9]Guangdong Provincial Key Laboratory of Advanced Thermoelectric Materials and Device Physics, Southern University of Science and Technology, Shenzhen, 518055, China

#Contributed equally

*Correspondence to qypcool@ustc.edu.cn, xielin@gbu.edu.cn, dxiang@sjtu.edu.cn, hejq@sustech.edu.cn.

Quantum many-body interactions and strong correlation effects facilitate the emergence of new quantum states and exotic properties in condensed matter systems. One of outstanding questions in this field is whether a non-ergodic and non-thermalizing state solely driven by strong correlations, i.e. the disorder-free many-body localization (DFMBL) phase, exist in condensed matter systems. In this work, using a novel Bragg scattering phase breaking approach and sophisticated theoretical simulation, we unambiguously identify this two-dimensional quantum state in the real material system of $AgCrSe_2$. The strong correlations among in-plane Ag atoms give rise to significant phononic flat bands and static two-dimensional off-center displacements of Ag atoms ranging from 0 to 0.5 Å with non-ergodic behavior. With a few representative materials, we prove that low-energy phononic flat bands serve as a promising indicator for a catalogue of DFMBL materials. As temperature rising, this static localization state with multiple degenerate local configurations persists up to near room temperature before transforming into a dynamically correlated state at a higher temperature. Emergent vibrational modes in the THz range and star-like topological structures constitute the low-energy excitation in the DFMBL phase. Moreover, the atomic engineering of a single Ag atom position induces a collective, correlated triangular motif, manifesting a generalized Winger solid in the atomic degree of freedom and underpinning the DFMBL regime. The identified DFMBL phase in a real material system offers a unified framework for understanding the rich quantum phenomena, such as the boson peak, the Debye-Einstein two-channel thermal conductivity and the spiral spin liquid state in quantum materials, and also offers a promising platform for topologically protected quantum computation and quantum memory.

Quantum many-body interactions and strong correlation effects have attracted broad interest because they give rise to exotic electric and thermal conductivity and underpin many quantum phases and phenomena in the field of condensed matter physics. These include topological order [1,2], high temperature superconductivity [3,4], Wigner molecules/solids [5,6], strange metals [7,8], fractional quantum Hall states [9,10], quantum spin liquids [11] and Debye-Einstein two-channel thermal conductivity [12]. In these quantum phenomena, Coulomb energy dominates over kinetic energy through geometry frustration and interaction-driven mechanisms, yielding flat bands in momentum space and strong static/dynamic localization of quasiparticles in real space [5,6,13-20]. Characterized by ergodicity breaking and non-thermalization behavior, such a strong correlations driven localization is commonly referred to as disorder-free many-body localization (DFMBL) [21,22], which is distinct from both Anderson localization and conventional many body localization driven by quenched disorder [13,23,24]. While the DFMBL phase has been predicted theoretically [21,25-27] and recently observed in synthetic quantum matters (cold atoms, trapped ions and superconducting qubits) [22,28-30], its experimental realization in a real solid-state material system has remained elusive. Center questions for the DFMBL phase are hence: Dose the DFMBL phase with multiple degenerate local configurations exist in real material systems? Is there a deterministic relationship between flat bands and the DFMBL phase? Is there a static to dynamic transition of the DFMBL phase? Is it possible to engineer the on-site localization in the DFMBL phase? The exploration of this exotic phase provides a unified scenario for understanding the quantum properties in condensed matter physics and for developing of topologically protected quantum computation and quantum memory.

In this study, we employ single-crystalline $AgCrSe_2$ as a model system because of the coexistence of phononic flat bands and a suit of anomalous properties associated with many-body localization in this material system. The phonon spectrum $AgCrSe_2$ by many-body perturbation theory displays significant low-frequency phononic flat bands (~ 0.7 THz) [31]. These phononic flat bands have been theoretically predicted to host extreme localizations in the atomic degree of freedom [32]. At low temperature (10-40 K), $AgCrSe_2$ displays an anomalous boson-peak-like feature [33] and unexpected vibration peaks spanning ~ 0.5-2 meV [34], indicating excessive vibrational modes in the THz range. This boson-peak-like feature, a distinctive many-body driven characteristic of amorphous materials [35,36], underpins a Debye-Einstein two-channel model for the ultralow thermal conductivity of $AgCrSe_2$ [33] and other thermoelectric materials [12,37], arising from localization in the atomic degree of freedom. $AgCrSe_2$ also has rich magnetic properties, such as the spiral spin liquid state [38] and the anomalous Hall effect [39]. These exotic magnetic behaviors always suffer complexity from the local crystal structure [38-42]. Therefore, the coexistence of phononic flat bands with exotic electric and thermal conductions renders $AgCrSe_2$ an ideal platform for identifying the DFMBL phase with multiple degenerate local configurations in the atomic degree of freedom.

To identify atomic-scale local correlated structures in material systems, we develop a novel method employing ultrafast structural response via femtosecond electron diffraction, which has been validated in our previous work [43,44]. In this study, we establish a systematic methodological framework for this approach and extend its application to the study of the DFMBL phase (see Fig. S1 and details in Supplementary Materials). Fig. 1A manifests the schematic representation of the femtosecond electron diffraction system (see details in Supplementary Materials). The femtosecond laser pulse serves as the external stimulation, and at varying time delays, another femtosecond electron pulse probes the diffraction pattern change, thereby extracting vibrational and displacive responses of the materials with ~ 50 femtosecond and milli-angstrom spatiotemporal resolution. In a real crystal structure with local correlated structures, atomic displacements induced by strong correlation are generally concealed within an average crystal structure. Such atomic displacements render the imaginary part of the structural factor non-zero and break the the linear $s^2$ (s is the scattering vector) dependence of intensity changes over characteristic Bragg peaks, which is termed a Bragg scattering phase breaking regime (see details in Supplementary Materials), as illustrated in Fig. 1B. Therefore, the disrupting of the $s^2$ dependence serves as a signature of local correlated structures deviating from the average symmetry and the local correlated structures in real space are then deduced from this anomalous intensity changes across characteristic Bragg peaks.

On average, $AgCrSe_2$ hosts a hexagonal crystal structure with alternative Ag layers and CrSe6 octahedral layers stacked along the c axis [45], as shown in Fig. 1C. At $T_c$ of about 450 K, the space group of crystal symmetry changes from R3m to R-3m. Both the two phases hold characteristic Bragg reflections (110), (220) and (330) with the imaginary part of the structural factor to be zero (see Table S1 and S2 in Supplementary Materials). In this work, using femtosecond electron diffraction and exploiting the novel Bragg scattering phase breaking regime, we reveal static, correlated two-dimensional displacements of Ag atoms away from their average (idealized) lattice positions in single-crystalline $AgCrSe_2$. The competition between Ag-Ag Coulomb correlations and potential wells induced by $CrSe_2$ layers, leading to phononic flat bands and driving the system into a DFMBL regime. The absence of such Cu displacements in the isostructural sister compound $CuCrSe_2$, as illustrated in Fig. 1C, further validates the deterministic relationship between phononic flat bands and local correlated structures. These unique local correlated structures featuring non--ergodic behavior, constitute the first experimental evidence of DFMBL with topological structures in a real material system. The single crystalline $AgCrSe_2$ and $CuCrSe_2$ sample with the thickness of ~ 50 nm is prepared and transferred onto a TEM copper grid for subsequent experiments (see sample preparation in Supplementary Materials). The sample is pumped with 400 nm femtosecond laser at the base temperature of 11-440 K. The ~50 fs temporal resolution of our femtosecond electron diffraction system [46] allows for the detection of the fastest structural response in most material systems, thereby fulfilling the experimental requirement efficiently.

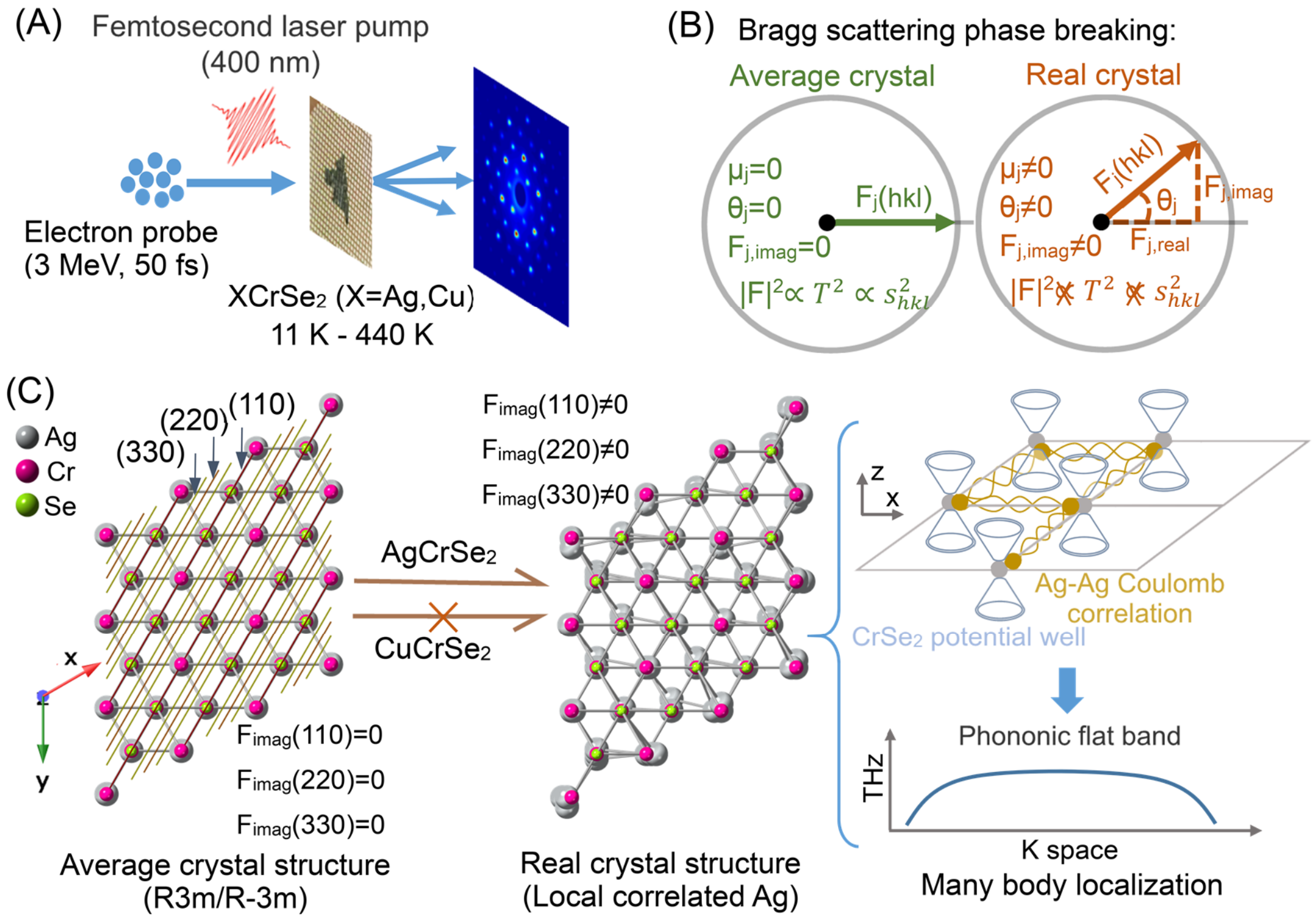

Fig. 1 (A) Schematic representation of the femtosecond electron diffraction system. (B) Schematic illustration of the identification of local correlated structures with Bragg scattering phase breaking regime. The complex form of the structure factor F(hkl) illustrated by a circular diagram. The amplitude $F_{real}$ is represented as the radius of the circle and the phase $F_{imag}$ as the angle made to the horizontal direction. The local displacement $\mu_j$ makes Bragg scattering phase $\theta_j \neq 0$, disrupting the $s^2$ dependence of the Bragg intensities. (C) Contrast between the average (idealized) crystal structure and the real crystal structure of $AgCrSe_2$, absent in $CuCrSe_2$. The right panel illustrates the competition between Ag-Ag Coulomb correlations and potential wells induced by $CrSe_2$ layers, leading to phononic flat bands and driving the system into a many body localization regime. The gray circles denote Ag atoms at the idealized positions centered at the $CrSe_2$ potential wells and the brown circles indicate real positions of Ag atoms determined by the competition between Ag-Ag Coulomb correlations and $CrSe_2$ potential wells.

The differential diffraction pattern at the time delay of ~ 6 ps and the time-resolved structural response of the several characteristic Bragg peaks at 100 K are shown in Fig.

2A and 2B, respectively. The monotonic intensity decay from femtosecond to picosecond timescales indicates enhanced thermal vibrations. Therefore, the intensity changes across these peaks are expected to follow a linear $s^2$ dependence under small perturbation. However, the experimental intensity change –In(I/I0) at the time delay of ~ 6 ps, as shown in Fig. 2C, deviates significantly from the linear dependence. For example, the intensity change of (220) is expected to four times larger than that of the (110), while in experimental results, the intensity of (110) decays by ~ 5% and the (220) show somewhat intensity enhancement. Such anomalous intensity changes arise since the onset of the structural response and persists as varying pump fluence, indicating that it is an intrinsic property of the material. The six reflections of the {220} crystal plane family labeled by green circles display the same anomalous intensity change, so the possible impact from sample tilting and crystalline domain can be neglected. At the much lower temperature of 11 K, the similar structural response deviating from linear dependence is shown in Fig. S2 in Supplementary Materials. Fig. S3 in Supplementary Materials presents differential diffraction patterns at distinct time delays. Possible contributions from multiple scattering, photoinduced displacements and local strains are excluded (see Supplementary Materials). All above analyses point to static atomic-scale local structures in $AgCrSe_2$ at low temperatures, which break the zero phase of these characteristic Bragg peaks and give rise to anomalous intensity changes. In contrast, the structural response at the elevated temperature of 440 K follows a quasi-linear $s^2$ dependence, as shown in Fig. 2D-2F. For instance, the intensity change of (220) is as expected to around four times larger than that of the (110). The high temperature of 440 K is close to the transition temperature of the R-3m phase. Both the low temperature R3m phase and high temperature R-3m phase exhibit zero phase for these characteristic Bragg peaks, so the linear $s^2$ dependence is expected for both phases. The structural transition from R3m to R-3m mainly derives from Ag displacements and the host sublattice CrSe2 remains stable across temperatures [45]. Therefore, we attribute the atomic-scale local structures at low temperature to local Ag displacements deviating from the idealized lattice positions in the average crystal structure. The multiple local correlated Ag displacements constitute the average Ag position in the R3m phase. As the temperature rising, the thermal fluctuation suppresses and overwhelms the local Ag displacement and then gives rise to the linear $s^2$ dependence as expected, so a static to dynamic transition of the local structures unifies the complete physical process. Moreover, as an isostructural sister compound of $AgCrSe_2$, $CuCrSe_2$ exhibits a quasi-linear $s^2$ dependence in its structural response at the temperature of 30 K (and 200 K), as shown in Fig. 2G-2I (and Fig. S4 in Supplementary Materials), indicating the absence of local correlated displacements. Subsequent DFT simulations reveal that phononic flat bands driving such local correlated displacements are a distinctive feature of $AgCrSe_2$ but are absent in $CuCrSe_2$.

We take the ratio between the intensity change of $I_{220}$ and $I_{110}$, i.e. $\Delta I_{220}/\Delta I_{110}$, as the parameter to evaluate the static to dynamic transition of the local structures in $AgCrSe_2$, as shown in Fig. 3A. As temperature rising, the ratio gradually changes from the negative values to the theoretically predicted value of ~ 4 with the ideal crystalline structure, unambiguously evidencing the temperature-driven static to dynamic transition of the local structures. At the fixed temperature of 330 K, we observe the ratio gradually changes from ~ 2 to the expected value of ~ 4 as increasing the pump fluence, therefore, the laser pump also modulates the local structures. Such a continuous transition with multiple intermediate states featured with local correlated structures is beyond the conventional framework of first/second order transition and the Landau transition theory [47,48]. Fig. 3B illustrates the physical scenario revealed in our study: a transition from the static local structures below 100 K to the dynamic local structures at 440 K. In the real material system of $AgCrSe_2$, at the temperature below 100 K, multiple local potential minima (e.g., p0, p1, p2 in Fig. 3B) instead of a single minimum allow Ag atoms at different crystal sites to occupy positions deviating from their ideal positions. These static Ag site deviations induce local vibrations and potentially excess vibrational density of states. At the high temperature above 440 K, the thermal fluctuation overwhelms the potential barriers among multiple local minima. As a result, the dynamic Ag deviations with random occupation of multiple local minima emerge, giving rise to the averaged occupation of the idealized crystal sites. In the range of 100 K ~ 440 K, the competition between the local correlated interaction and the thermal fluctuation leads to multiple intermediate states. Our recent work reveals that the thermal fluctuations of Ag atoms at the high temperature are nontrivial, exhibiting a pronounced diffuse ring pattern as shown in Fig. 3C. This diffuse ring feature represents dynamically evolving Ag-Ag trimers driven by Coulomb interactions [49], which indicates a dynamical localization phase with non-thermalization behavior in this many body system.

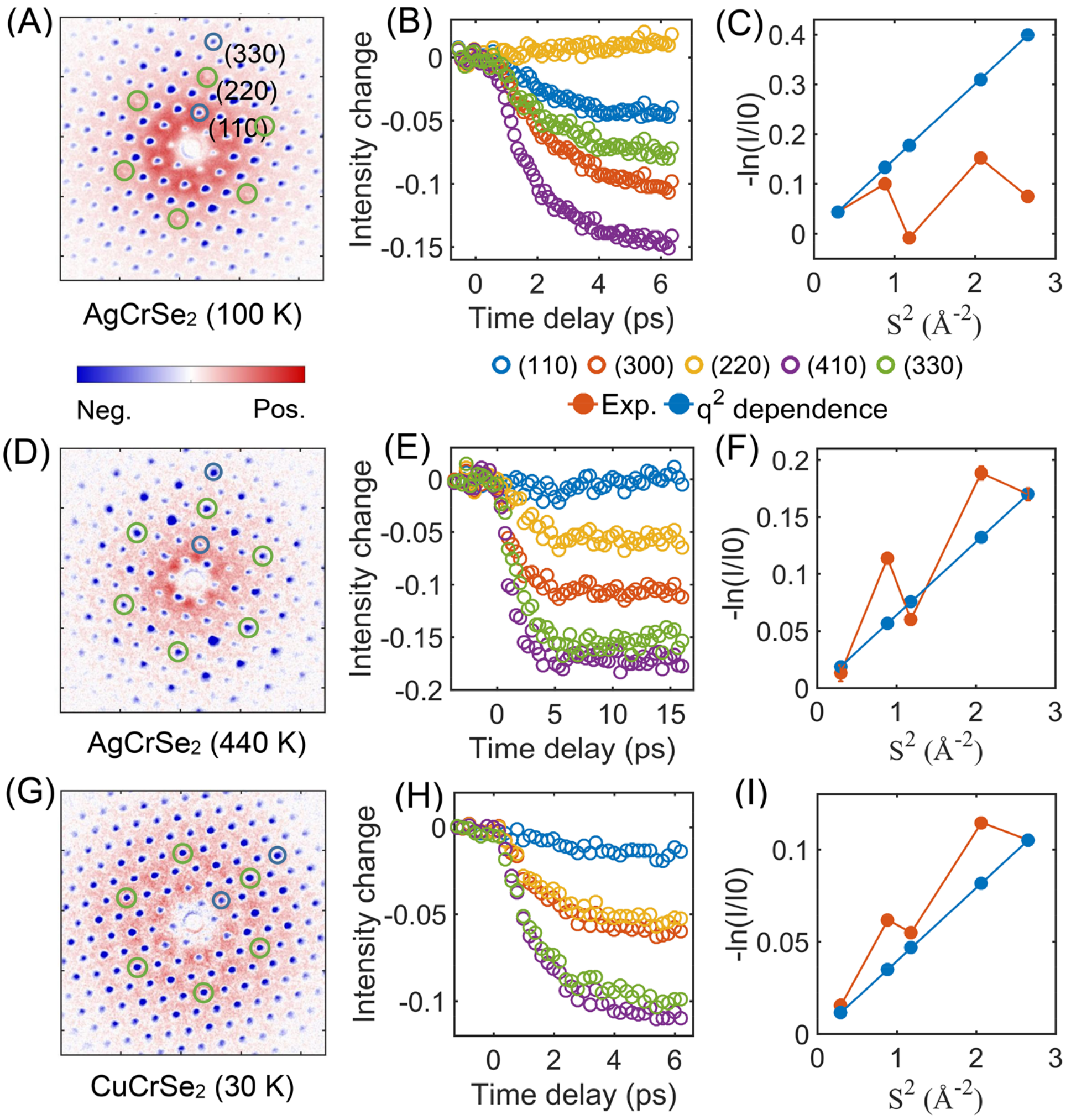

Fig. 2 Ultrafast dynamics of AgCrSe2 measured at base temperature of 100 K (A-C) and 440 K (D-F), and of CuCrSe2 measured at a base temperature of 30 K (G-I). Pump fluences are 3.4 mJ/cm$^2$ (A-F) and 2.6 mJ/cm$^2$ (G-I). (A) Differential diffraction pattern at the time delay of ~ 6 ps. The circles label the characteristic Bragg peaks (110), (220) and (330). (B) Experimental intensity changes of characteristic Bragg peaks at 100 K. (C) The $s^2$ (i.e., $q^2$) dependences of the experimental results at 100 K and the theoretical prediction results. (D) Differential diffraction pattern at the time delay of ~ 6 ps. (E) Experimental intensity changes of characteristic Bragg peaks at 440 K; (F) The $s^2$ dependences of the experimental results at 440 K and the theoretical prediction results. (G) Differential diffraction pattern at the time delay of ~ 6 ps. (H) Experimental intensity changes of characteristic Bragg peaks at 30 K; (I) The $s^2$ dependences of the experimental results at 30 K and the theoretical prediction results. (J) Schematic illustration of many-body localization modulation via temperature-dependent suppression (D-F) and phononic-flat-band dependent suppression (G-I).

To quantify the atomic-scale local structures, we perform structure factor calculations by incorporating local Ag displacements to reproduce the experimental anomalous intensity changes observed at the low temperature. To simplify the calculation, we use a single unit cell, rather than a supercell, to model the diffraction intensity change of the overall lattice. Nine parameters are utilized to constrain the local structure model: the displacements along the x and y axes for the three Ag atoms in the unit cell, as well as the three vibrational amplitudes for Ag, Cr and Se atoms respectively (see calculation details in Supplementary Materials). 48 local structures and their corresponding vibration amplitudes are identified as shown in Table S3 in Supplementary Materials. Fig. 3D displays the contrast between the unit cell in the ideal R3m phase and the identified local structure with local Ag displacements (i.e., the first structure listed in Table S3 in Supplementary Materials). The local Ag displacement as large as 0.37 Å is widely observed across these local structures. To simulate the time–resolved intensity change, the thermal vibrations gradually increase to the identified value with the same time constant as that of the experimental results. With the local structure shown in Fig. 3D (bottom), the simulated intensity changes of the characteristic Bragg peaks as a function of the time delay manifest in Fig. 3D (top). These simulation results well reproduce the experimental results in Fig. 2B. Therefore, the structure factor calculation quantitatively identifies the static local structures with local Ag displacements in single-crystalline $AgCrSe_2$. As we expand the boundary of the amplitude of the Ag displacements to ~ 0.5 Å, more than a thousand local configurations enable to reproduce the experimental results (see Table S4 in Supplementary Materials). The finite number of the local figurations indicates that the local displacements among Ag atoms are correlated and non-ergodic instead of random. The present calculation is based on a single unit cell, so the identified local structures represent an average of the local structures in real materials. A more precise structural model with a supercell is expected and demands much computational source, which is beyond the scope of the present work.

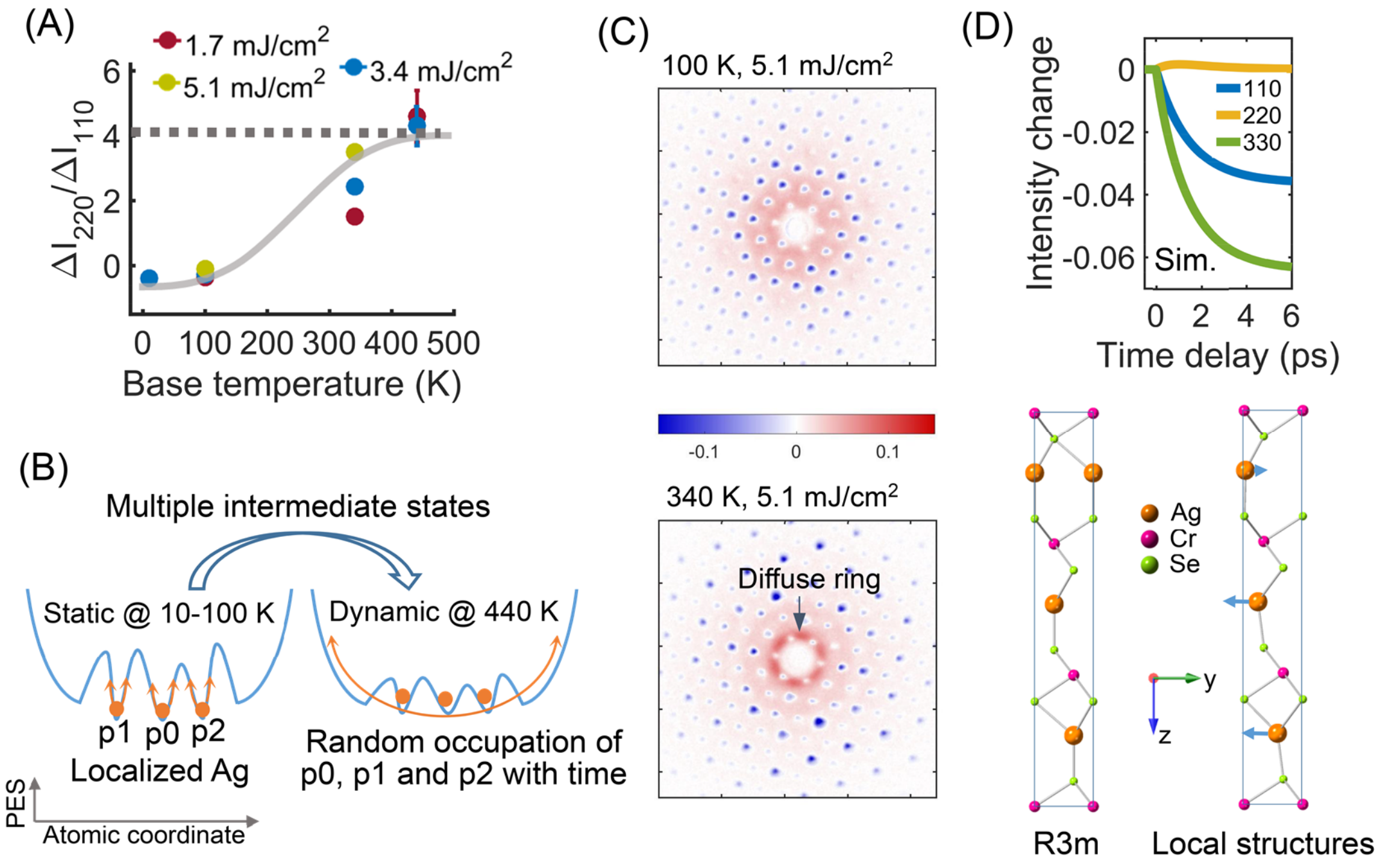

Fig. 3 (A) $\Delta I_{220}/\Delta I_{110}$ as a function of the base temperature and pump fluence. The solid line is a guide to the eye and the dotted line is the theoretically predicted ratio from the ideal crystalline structure. (B) Schematic representation of the static to dynamic transition of the local structures from the perspective of potential energy landscape. (C) Differential diffraction pattern at the time delay of ~ 9 ps. An arrow indicates the diffuse ring feature at the high temperature. (D) (top) Simulated intensity change as a function of time with local correlated structures. (bottom) Rigid unit cell in the R3m phase (left) and the local correlated structure with Ag displacements indicated by arrows (right).

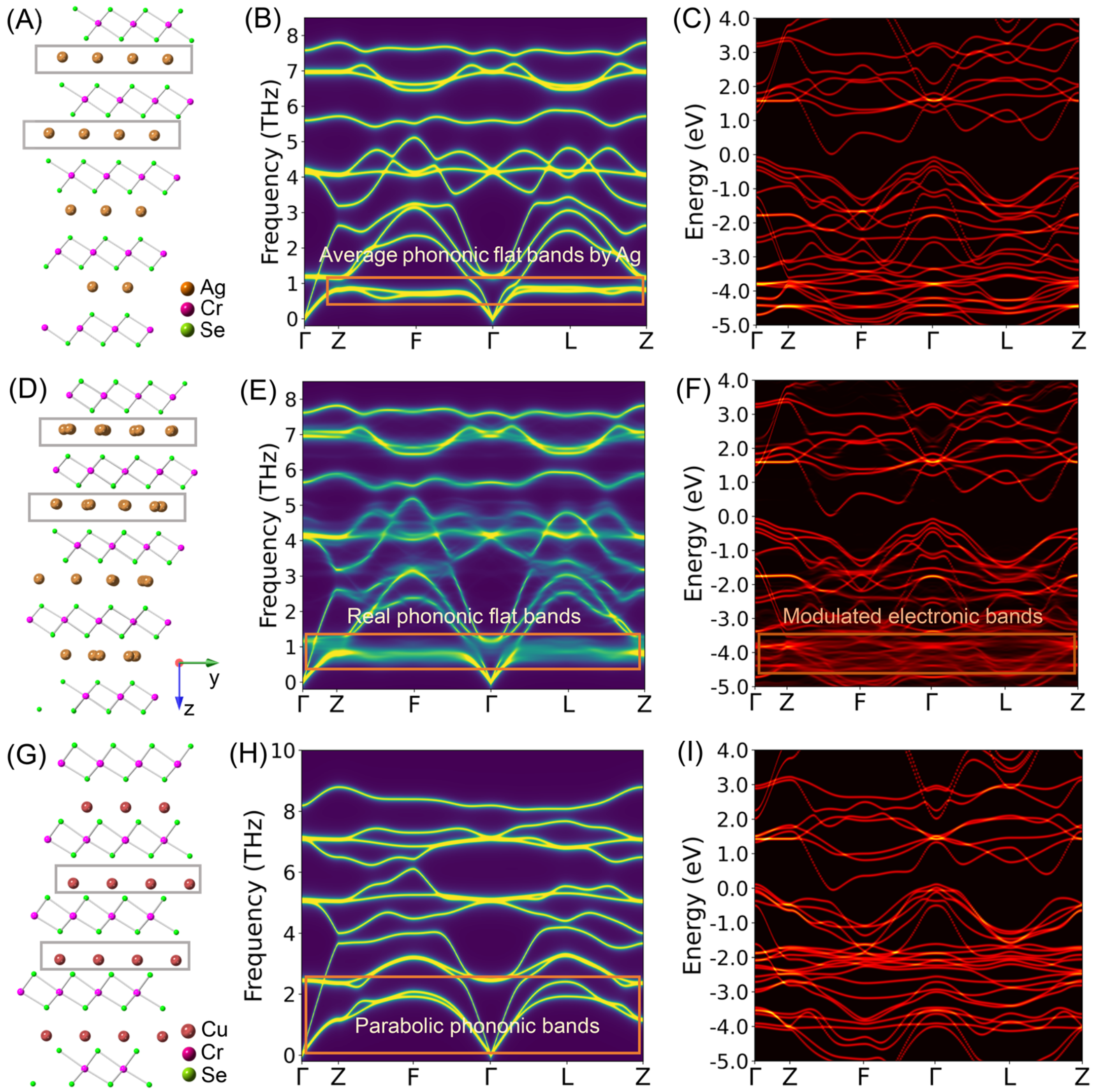

Fig. 4 DFT simulation of the average high-symmetry R3m structure of $AgCrSe_2$ (A-C), the real R3m structure of $AgCrSe_2$ featuring local correlated Ag displacements (D-F), and the real R3m structure of $CuCrSe_2$ lacking such local correlated displacement (G-I). (A) Average (idealized) R3m crystal structure with Ag atoms rigidly positioned at the centers of Se tetrahedra. (B) Phonon spectral function along high symmetry k-path. The highlighted region (rectangle) reveals phononic flat bands predominantly of Ag character. (C) Electron band structure along the high symmetry k-path. (D) Realistic crystal structure with local correlated Ag displacements. (E) Phonon spectral function along the high symmetry k-path. The rectangle marks the real phononic flat bands. (F) Unfolded electron band structure along the high symmetry k-path. The spectrum in the rectangle indicates the modulated electronic bands by local Ag displacements. (G) Realistic crystal structure with Cu atoms rigidly positioned at the centers of Se tetrahedra. (H) Phonon spectral function along the high symmetry k-path. The rectangle marks the parabolic phononic bands. (I) Unfolded electron band structure along the high symmetry k-path.

To further quantify the local correlated structures in the low temperature phase of $AgCrSe_2$, we perform Density-functional theory (DFT) simulations. In the framework of density-functional perturbation theory, the supercell with 256 atoms is stable in the average R3m phase and the Ag atoms uniformly located at high symmetry positions, as shown in Fig. 4A. The phonon spectral function in Fig. 4B manifests remarkable phononic flat band at around 0.7 THz and the Ag atoms dominate these low-frequency flat bands (Fig. S5 in Supplementary Materials). The electron band structure of the idealized R3m phase is shown in Fig. 4C and the Ag atoms dominate the deep level electronic band around – 4 eV. By applying anharmonic special displacements to the supercell and then allowing a full relaxation of all the ion positions (see details in Supplementary Materials), the relaxed supercell in Fig. 4D displays remarkable two-dimensional local Ag displacements deviating from the ideal sites, meanwhile, the positions of Cr and Se atoms remain unchanged. The energy for the supercell with local correlated structures is lower than that of the ideal structure by ~ 1 meV/atom, therefore, these local structures are energetically favorable. Most Ag atoms host significant in-plane local displacements and the maximum amplitude is ~ 0.5 Å, which is comparable to the value of 0.37 Å by structure factor calculation in Fig. 3C. In Fig. 4E, the phonon spectral function of the supercell with local structures do not show any signature of imaginary frequency, indicating that the supercell is dynamically stable. The real crystal structure with local displacements retains the same distinctive phononic flat band at around 0.7 THz. In the electron spectral function in Fig. 4F, excessive band structures emerge at around – 4 eV and give rise to a modulated quasi-flat electronic bands. The Ag atoms dominate these deep-level electronic band structures, so the quasi-flat band feature is closely associated with the anomalous local structures. Phononic flat bands have been theoretically predicted to host extreme localizations in the lattice freedom [32]. To the best of our knowledge, the localized Ag displacement featured with phononic flat bands is the first experimental examplification of the strong correlation driven many-body localization in a real material system, i.e., a novel DFMBL phase. Such localization induced by anharmonic special displacements has a purely dynamical origin, which agrees well with the quantum model [21]. As a sister compound of $AgCrSe_2$, $CuCrSe_2$ displays a conventional parabolic band structure in its low frequency phonon dispersion (Fig. 4H). With identical structural relaxation conditions as those applied in Fig. 4D-4F, $CuCrSe_2$ shows no local displacements of Cu atoms (Fig. 4G). In our femtosecond electron diffraction measurements on $CuCrSe_2$ (Fig. 2G-2I), the intensity change of the characteristic Bragg peaks at 30 K follows a linear $s^2$ dependence, suggesting the absence of Cu local displacements. Taken together, these observations establish a direct and deterministic correlation between the DFMBL phase and the phononic flat bands. Specifically, the competition between Ag-Ag Coulomb correlations and potential wells induced by $CrSe_2$ layers, leading to the phononic flat bands and driving the system into a DFMBL phase.

To further characterize this novel DFMBL phase, we perform macroscale machine-learning molecular dynamics calculations using a supercell with 17280 atoms (see details in Supplementary Materials). Fig. 5A presents the color mapping of Ag-Ag bond lengths within a single Ag layer of the $AgCrSe_2$ supercell. In contrast to the idealized $AgCrSe_2$ structure with a uniform bond length of 3.68 Å, the irregular distribution of bond lengths in Fig. 5C suggests non-ergodic and non-thermalization behavior, a hallmark of the DFMBL phase. The two-dimensional nature of the localization displays in Fig. S6 in Supplementary Materials. Within a bond length of 3.48 Å, most Ag atoms arrange into triangular and rhombic motifs, as shown in Fig. 5B, suggesting attractive Coulomb interaction and star-like topological character. The inelastic neutron scattering experiments in the left figure of Fig. 5B display two distinct low-energy peaks within a 2 meV scale in $AgCrSe_2$ [34,50], which are absent in $CuCrSe_2$. Crucially, these two peaks vanish upon heating to 300 K, signaling pronounced lattice anharmonicity [34,50]. Their emergence and temperature-dependent suppression suggest that they arise from emergent vibrational modes induced by the novel DFMBL phase in $AgCrSe_2$. Based on the calculated macroscopic strcuture in Fig. 5A, we compute the static structure factor for neutron scattering (see details in Supplementary Materials). As depicted in Fig. 5B, the calculation results reproduce the two distinct peaks within the 2 meV energy window; moreover, these two peaks get weakened or vanished if calculating with the average R3m phase. This stark contrast agrees with the emergent nature of these two vibrational modes induced by DFMBL.

Fig. 5E displays the atomic engineering of the DFMBL phase by a writing-reading regime. With DFT simulation, a single Ag atom in the ideal R3m phase is vertically moved of 0.6 Å by a local tweezer (writing), like the electron beam in a TEM [51], as show in Fig. 5E (left). The supercell is first relaxed with fixed atomic positions to mimic the experimental condition of a local tweezer [51], then further relaxed without any constraints. In this case, a triangular motif with shorter bond lengths by 0.1-0.2 Å displays in Fig. 5E (right) and exhibits the same topological structure as that in Fig. 5A and 5B. This energetically favorable triangular motif by non-thermal local manipulation thereby ambiguously demonstrates the Ag-Ag Coulomb correlation in the DFMBL phase. Analogous to the triangular Wigner molecule charge configuration in a moiré superlattice [5], the triangular motifs in Fig. 5A, 5B and 5E thereby manifest a generalized Winger solid in the atomic degree of freedom. Simulations of movement along the horizontal direction further confirm the triangular motif and the strong correlation in the DFMBL phase (see Fig. S7 in Supplementary Materials). The modulation of the local electron/spin properties induced by these elegant atomic engineering serve as a readout signal. Therefore, a complete atomic engineered writing-reading regime based on strong correlation, beyond vacancy-interstitial defect complexes [51], in real material systems is proposed.

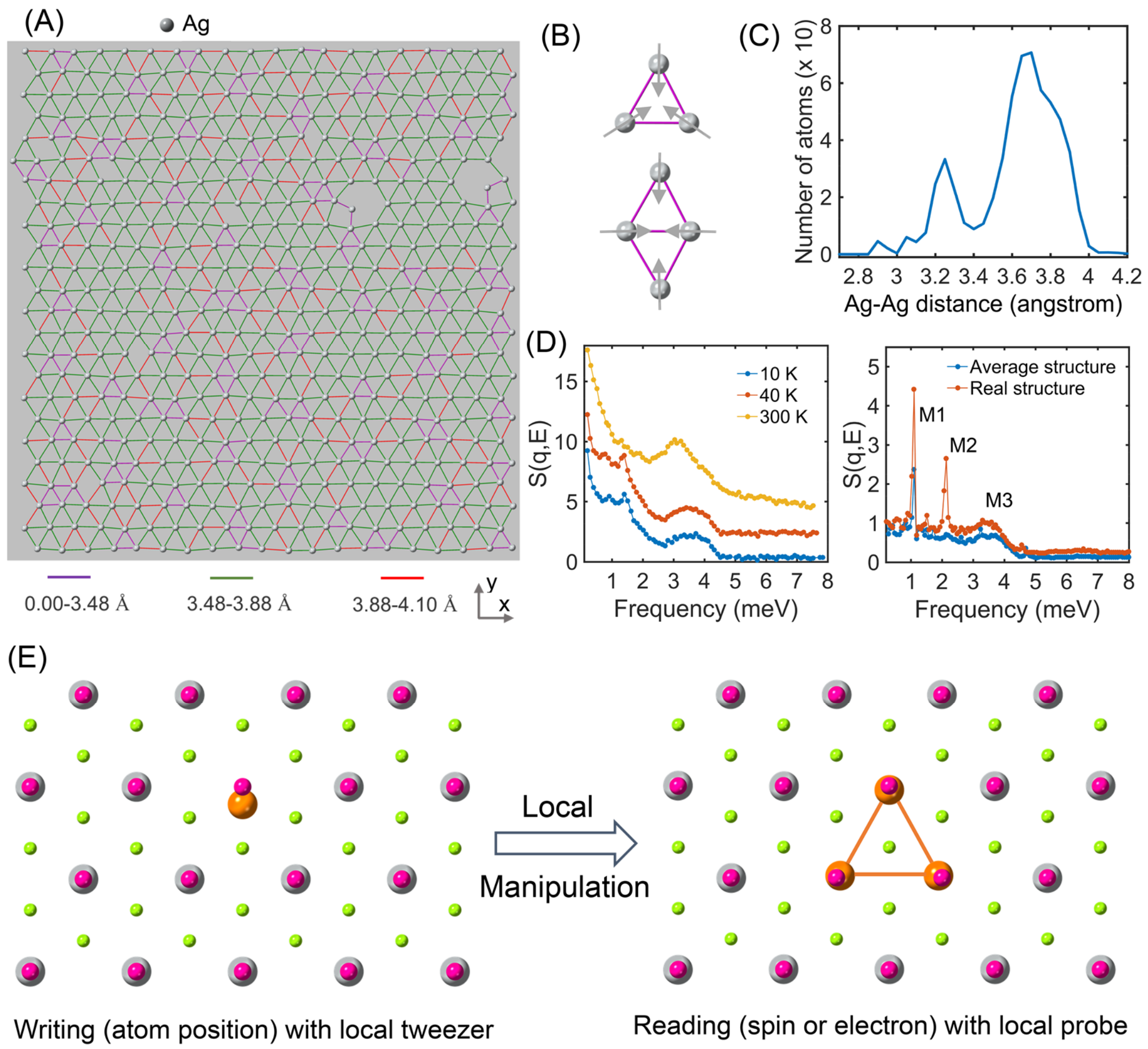

Fig. 5 Characterization of the DFMBL phase in $AgCrSe_2$. (A) Macroscopic visualization of many body localization, encoded by color mapping of Ag-Ag bond lengths. (B) Triangular and rhombic motifs enclosed with shorter bond lengths (≤ 3.48Å). The arrows indicates the movement directions of atoms. (C) Distribution of the Ag-Ag bond lengths. (D) Inelastic neutron scattering (left panel, from the literature) and DFT simulation (right panel) revealing the emergence of two distinct vibrational modes (M1 and M2), driven by many body localization. (E) DFT simulation of the atomic engineering of the DFMBL phase with a writing-reading regime. The left panel is the engineering of a single Ag atom position with local tweezer. The right panel is, after relaxation, a triangular motif with shorter bond lengths induced by strong correlation, which can be read using an electron or spin probe.

In this study, with the combination of the femtosecond electron diffraction, the novel Bragg scattering phase breaking regime, the sophisticated structure factor calculation and the state-of-the-art density functional theory simulation, we unambiguously reveal the novel DFMBL phase in the real material system of $AgCrSe_2$. The competition between the two-dimensional Ag-Ag Coulomb correlations and potential wells induced by $CrSe_2$ layers, leading to phononic flat bands and driving the system into a DFMBL regime. We demonstrate the deterministic relationship between the phononic flat bands and the DFMBL phase, therefore, a catalogue of naturally occuring DFMBL materials with low-energy phononic flat bands is promising, analogous to the catalogue of electronic flat bands [19]. Several representative material systems are discussed to elucidate the roles of imaginary phononic flat bands and energy-dependent phononic flat bands in determining DFMBL materials (see Supplementary Materials for details). In our recent study [52], robust sub-angstrom localization within the sub-unit cell of $BaTiS_3$ provides evidence for a DFMBL phase driven by imaginary phononic flat bands. As the temperature rises, the static DFMBL phase transitions to a dynamic DFMBL phase, both of which exhibit non-ergodic and non-thermalizaiton behavior consistent with the recent dynamical localization model with bosonic gas [16]. Such a static to dynamic transition while retaining the strong correlation is analogous to the Bardeen-Cooper-Schrieffer (BCS) to Bose-Einstein condensate (BEC) crossover in flatband superconductivity [53], yielding an interesting topic regarding the mechanism of the superconductivity. We also demonstrate the atomic engineering of the DFMBL phase by a writing-reading regime. These results on the novel DFMBL phase offer a promising platform for topologically protected quantum computation and quantum memory [51,54,55].

In the DFMBL phase of $AgCrSe_2$, the two-dimension sub-angstrom localization and the emergent vibrational modes in the THz range (below ~ 2 meV) constitute the microscopic origin of the boson-peak-like feature and the Debye-Einstein two-channel model (i.e., a phonon conduction channel and a hoping channel of localized vibrations) for the ultralow thermal conductivity of $AgCrSe_2$ [33] and other thermoelectric materials [12,37]. The emergent low-energy vibrational modes likely correspond to fractional excitations in the atomic degree of freedom, analogous to the fractional electronic excitation [56,57]. By superexchange interactions [58], the multiple local correlated structures indicated by Ag displacements probably break the interlayer antiferromagnetic state and induce the spiral spin liquid state featured by extensive degenerate ground states and strong short-range correlations [38,40]. Moreover, these local correlated structures constitute a structural origin of the time-reversal symmetry breaking responsible for the anomalous Hall effect in $AgCrSe_2$ [39,41], which is distinct conventional intrinsic (Berry-phase curvatures) and extrinsic (disorder scattering) mechanisms.

Characterizing the DFMBL phase in crystalline materials is actually entering a realm where the concept of constructing a three dimensional crystal with a unit cell is only partly valid [59,60]. Alternatively, the average over a multitude of local structures constructs what is called the average crystal structure. In this study, we propose the Bragg scattering phase breaking regime to unambiguously reveal the sub-angstrom local structures deviating from the average crystal structure in crystalline materials. This novel regime has an intrinsic advantage in distinguishing the static displacements from dynamic vibrations, which is significant in constructing structure-property relationship but fails in other probing techniques. Instead of photoinduced structural dynamics [61,62], we extend the femtosecond electron diffraction to study the intrinsic anomalous structural characteristic in the ground state in quantum materials. Plenty materials exhibit characteristic Bragg peaks as summarized in Table S6 in Supplementary Materials, so the Bragg scattering phase breaking regime, combined with femtosecond electron diffraction, provides a powerful approach for uncovering hidden local structures and bridging the structure-property relationship from the innovative perspective of atomic-scale local structures in condensed matter physics and material science.

## Acknowledgements

**Funding:** This work was supported by National Natural Science Foundation of China (12304024, 12574045, 12525501, 12335010) and National Key R&D Program of China (no. 2024YFA1612204). D.C. acknowledges the support from National Natural Science Foundation of China (Grant Numbers 12574348) and Natural Science Foundation of Anhui Province (Grant Numbers 2408085MF176). D. X. would like to acknowledge the support from the New Cornerstone Science Foundation through the Xplorer Prize. The UED experiment was supported by the Shanghai soft x-ray free electron laser facility. J. H. acknowledges the support from the National Key Research and Development Program of China (2024YFB3813800), the National Natural Science Foundation of China (Grant No. 12434001, 52461160258, U25A20239), Guangdong

Provincial Key Laboratory of Advanced Thermoelectric Materials and Device Physics (Grant No. 2024B1212010001), the Outstanding Talents Training Fund in Shenzhen (Grant No. 202108) and 2024JC08A027. **Author Contributions:** Yingpeng Qi devised the project and conceived the presented ideas. Jianmin Yang prepared and characterized the sample. Yingpeng Qi performed the MeV femtosecond electron diffraction experiments with help from Zhihui Zhou, Qing Xu, Yang Lv, Xiao Zou, Tao Jiang and Pengfei Zhu. Yingpeng Qi analyzed the experimental data and built up the model presented in the paper. Lin Xie performed the DFT simulation. Yingpeng Qi wrote the paper with contributions from all other authors. Yingpeng Qi, Lin Xie, Dongxue Chen, Zhenrong Sun, Dao Xiang and Jiaqing He supervised the project. **Competing interests:** The authors declare no competing interest. **Data and materials availability:** All data needed to evaluate the conclusions in the paper are present in the paper and/or the Supplementary Materials. Additional data related to this paper may be requested from the corresponding author Yingpeng Qi.

# Disorder-free many-body localization in real quantum materials

Yingpeng Qi[1,2#*], Jianmin Yang[3,4#], Zhihui Zhou[2], Qing Xu[2], Yang Lv[2], Xiao Zou[2], Tao Jiang[2], Pengfei Zhu[2], Dongxue Chen[1,5], Zhenrong Sun[6], Lin Xie[7*], Dao Xiang[2,8*], Jiaqing He[3*]

[1]Suzhou Institute for Advanced Research, University of Science and Technology of China, Suzhou 215123, China

[2]Key Laboratory for Laser Plasmas (Ministry of Education), School of Physics and Astronomy, Shanghai Jiao Tong University, Shanghai 200240, China.

[3]Department of Physics, State Key Laboratory of Quantum Functional Materials, and Guangdong Basic Research Center of Excellence for Quantum Science, Southern University of Science and Technology, Shenzhen 518055, China.

[4]Department of Materials Science and Engineering, National University of Singapore, 117575, Singapore.

[5]Department of Physics, University of Science and Technology of China, Hefei, Anhui 230026, China

[6]State Key Laboratory of Precision Spectroscopy, and School of Physics and Electronic Science, East China Normal University, Shanghai 200241, China

[7]School of Physical Sciences, Great Bay University, Dongguan 523000, China.

[8]Zhangjiang Institute for Advanced Study, Shanghai Jiao Tong University, Shanghai 201210, China.

#Contributed equally

*Correspondence to qypcool@ustc.edu.cn, xielin@gbu.edu.cn, dxiang@sjtu.edu.cn,

hejq@sustech.edu.cn.

## Sample preparation

Single crystal $AgCrSe_2$ was grown by the chemical vapor transport method. The samples for the UED experiments were first sliced parallel to the a-b plane of the crystal using a diamond knife mounted on an RMC ultramicrotome, and then placed onto holey carbon-coated copper grids. The thickness of the sample used in the experiment is ~50 nm with a lateral dimensions of ~300 μm. The high quality of the sample is well calibrated with transmission electron microscopy and transport measurements (ionic conductivity and heat capacity (see our recent work of Ref. 53)). The single crystal nanofilm of $CuCrSe_2$ with the thickness of ~ 60 nm was prepared in the same way. The transport measurement of the thermal conductivity and ionic conductivity evidences the good quality of the single crystal samples.

## MeV femtosecond electron diffraction system

For the MeV femtosecond electron diffraction system, the 800 nm, 30 fs and 200 Hz laser pulse (Vitara and Legend Elite Duo HE, Coherent) is split into the pump and the probe pulse. The pump pulse excites the sample and the probe pulse is frequency tripled in nonlinear crystals before illuminating a photocathode for electron pulse generation. After being accelerated by an intense radio-frequency field to relativistic velocity (~0.989c), the electron pulse goes through a double-bend achromatic lens for pulse compression and jitter removal. The spot size of the electron pulse on the sample is ~200 μm FWHM, around five times smaller than the size of the pump laser, ensuring a homogeneous photoexcitation. The diffraction pattern is imaged by a phosphor screen (P43) and recorded by an electron-multiplying charge-coupled device (Andor iXon Ultra 888). The overall temporal resolution of the MeV FED system is ~50 fs. Further details for the system can be found in Ref. 46 in the main text.

## Identification of atomic-scale local correlated structures by femtosecond electron diffraction

The structure of materials fundamentally determines their properties, which is a traditional paradigm across various fields in condensed matter physics and materials science [1-4]. Resolving the structure-property relationship is therefore pervasive throughout the research field [5-7], spanning crystalline to amorphous material systems. For crystalline materials, symmetry analysis based on ideal/average crystal structure and mean-field appropriation has been well-established for quantifying material properties both experimentally and theoretically [3,8-10]. However, defects, fluctuations, degenerate states and correlated interactions introduce complexity into the crystal structure, leading to atomic-scale local correlated structures that deviate from the long-range average structure [11-25]. These local correlated structures, with correlation lengths limited to sub-unit cell to a few unit cells, underpin many exotic properties incompatible with average symmetry and mean-field theory. Significant efforts have been devoted to characterize theses atomic-scale local structures and bridge the relationship between the local structures and exotic properties through experimental and theoretical approaches [5,26,27], such as Raman and X-ray fine absorption spectrum [12,28], scanning transmission electron microscopy enhanced by computational techniques [29,30], X-ray and neutron diffraction [27,31] and computational predictions enhanced by machine learning models [5,32,33]. The lack of the statistic feature of the local probe [29,30] and the difficulty of retrieve local structure from halo-like diffuse scattering [27,31] urge novel methods with sub-angstrom spatial resolution. Moreover, all present approaches acquire the information of the local structures at equilibrium states; neither the dynamical response of the local structures nor the identification of the local structure with dynamical response has been achieved. As a result, elucidating the structure-property relationship from the perspective of atomic-scale local correlated structures remains an appealing yet unattained goal.

In this work, we establish a systematic methodological framework that employs ultrafast structural response via femtosecond electron diffraction to identify atomic-scale local correlated structures. This novel approach has been validated in our previous work [34,35], but has not been well established until now. As depicted in Fig. S1, defects, fluctuations and correlated interactions introduce atomic-scale local correlated structures, either static or dynamic, rendering the crystal less rigid than its average symmetry would suggest. Under external stimulation, the structural response is categorized into vibrational and displacive responses. Consequently, two regimes are introduced to identify local correlated structures: the Bragg scattering phase breaking [35], which uses the vibrational response, and the local symmetry breaking [34], which uses the displacive response. For femtosecond electron diffraction, the femtosecond laser pulse serves as the external stimulation, and at varying time delays, the femtosecond electron pulse probes the diffraction pattern change, thereby extracting vibrational and displacive responses of the materials. The intensity of the Bragg reflection is proportional to the square of the structure factor: $I_{hkl} \propto |F_{hkl}|^2$. The structure factor of the Bragg reflection (hkl) at the time delay t is expressed as follows:

$$F(hkl, t) = \sum_j T_j(t) \cdot F_j(t) = \sum_j T_j(t) \cdot f_j \cdot \exp\left(-i \cdot 2\pi \cdot s_{hkl} \cdot (r_j + \mu_j + \kappa_j(t)\right) \quad (1)$$

$$T_j(t) = \exp\left(-8\pi^2 \langle \nu_j(t)^2 \rangle s_{hkl}^2\right) \quad (2)$$

where the summation runs over all atoms within the local structures, $f_j$ denotes the atomic form factor for the jth atom, $T_j$ represents the thermal vibrations, $s_{hkl}$ is the scattering vector, $r_j$ indicates the vector position of the jth atom, $\mu_j$ is the intrinsic static local displacement induced by local correlated structure, $\kappa_j(t)$ signifies the photoinduced displacement and $\nu_j$ denotes the amplitude of the thermal vibration. For each atom, the structure factor $F_j$ is a complex number, composed of $F_{j,real}$, $F_{j,imag}$ and the phase angle $\theta_j$.

For the case of Bragg scattering phase breaking regime [35], in an average crystal structure with $\mu_j = 0$, some special characteristic Bragg peaks (hkl) exhibit the property that the phase (and the imaginary part of the structural factor) for each atom is zero, i.e. $\theta_j = 0$. This unique phase characteristic causes the intensity changes of these peaks to follow an $s^2$ dependence ($|F|^2 \propto s^2$, $\ln(|F|^2) \propto s^2$) as the amplitude of thermal vibrations increases within a relatively small range and/or photoinduced displacement $\kappa_j(t)$ with relatively small amplitude gets involved. This small amplitude approximation for vibrations and displacements is generally effective because even reaching the Lindemann's melting criterion, the square root of the particle mean-squared displacement from the equilibrium position is just ~ 0.1 of the interparticle distance. However, in real crystal structures, local atomic displacements ($\mu_j \neq 0$) are probably introduced in the ground state induced by defects, fluctuations, degenerate states and correlated interactions. In this case, the phase and the imaginary part become nonzero, i.e. $\theta_j \neq 0$, giving rise to the Bragg scattering phase breaking regime and the anomalous intensity changes disrupting the $s^2$ dependence. This is termed the Bragg scattering phase breaking regime, as illustrated in Fig. 1B. Therefore, the disrupting of the $s^2$ dependence serves as a signature of local structures deviating from the average symmetry and reversely, the local correlated structures can be extracted from the anomalous intensity changes across the characteristic Bragg peaks. The transverse coherence length of femtosecond electron pulse is a few nanometers, so it is sensitive to the local displacement ranging from sub-unit cell to a few unit cells. Note that, both vibrational response and displacive response by external stimulation contribute to the anomalous intensity changes across the characteristic Bragg peaks. With femtosecond electron diffraction, the temporal separation of these two intertwined physical processes-namely, the much faster displacive response compared to the vibrational response [34,36], enables the precise quantification of local structures. Other probable physical response, such as the relaxation of local strain [37], can also be identified by ultrafast structural dynamics and used to construct the local correlated structures. Any measurement at equilibrium state will suffer complexity and entanglement of these physical responses and obstacle quantifying local correlated structures.

For the case of local symmetry breaking regime [34], local correlated structures may induce a transformation of the potential energy surface from the single well of the average symmetry to a double well, as shown in Fig. S1. Consequently, under external stimulation, the instability of the double well potential energy surface leads to ultrafast local displacements [34]. These unexpected ultrafast local displacements represent another signature of local correlated structures deviating from the rigid average symmetry.

Therefore, the Bragg scattering phase breaking regime and local symmetry breaking regime compose a systematic methodological framework to identify atomic-scale local correlated structures and construct the structure-property relationship in quantum materials. In this study, we establish a systematic methodological framework for this novel approach and extend its application to the study of the DFMBL phase.

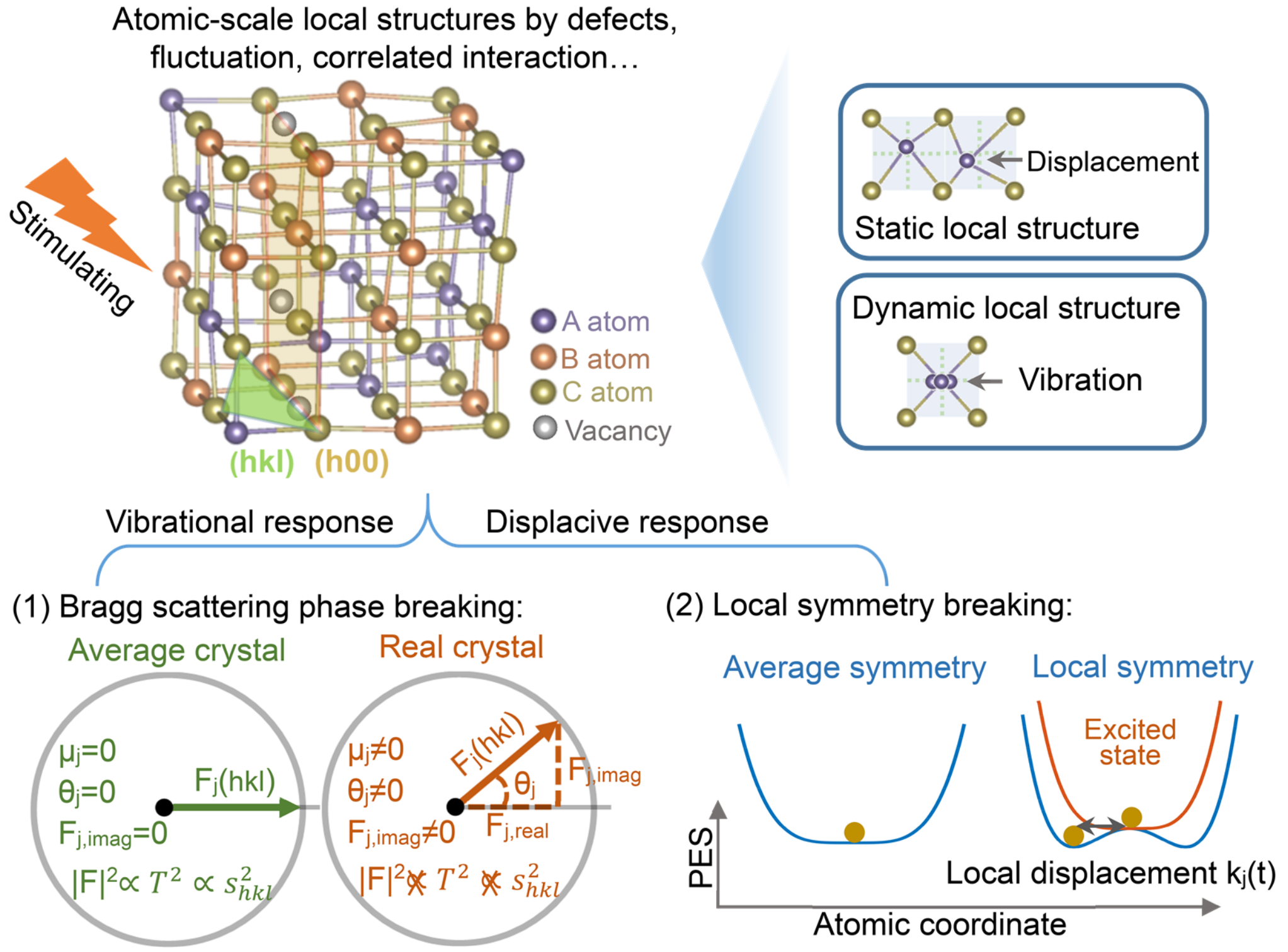

Fig. S1 Schematic illustration of the identification of local correlated structures, either static or dynamic, using (1) Bragg scattering phase breaking and (2) local symmetry breaking. For Bragg scattering phase breaking, the local displacement $\mu_j$ makes Bragg scattering phase $\theta_j \neq 0$, disrupting the $s^2$ dependence of the Bragg intensities. For local symmetry breaking, the instability of the local symmetry is depicted as a double well potential; external stimulation induces an excited state characterized by a local atomic displacement $k_j(t)$, resulting in a transformation of the potential energy surface from a double well to a single well.

## Ruling out multiple scattering effect, photoinduced displacements and local strains

Calculating the extinction distance $\xi$ of the (hkl) plane: $\xi = \pi V\cos(\theta_B)/\lambda F_{hkl}$, where V is the volume of the unit cell, $\theta_B$ is the Bragg angle, $\lambda$ is the electron wavelength and $F_{hkl}$ is the structure factor. The above formula suggests a higher sensitivity of the multiple scattering effect for the low-order Bragg reflection than the high-order reflections, therefore, we choose the (110) reflection to evaluate the extinction distance of $AgCrSe_2$. Here, V is 248.87 Å$^3$, $\cos(\theta_B) \simeq 1$ as we are working with small $\theta_B$, $\lambda$ = 0.00356 Å, $F_{110}$ = 39.61 Å, then the extinction distance $\xi$ = 554 nm, which is 12 times larger than the sample thickness of 46.7 nm used in the experiment. The multiple scattering effect can be neglected in this case. In addition, as a comparison, we perform femtosecond electron diffraction experiments on $CuCrSe_2$ with the thickness of ~ 60 nm, the isostructural sister compound of $AgCrSe_2$. The experimental results at the base temperature of 30 K in Fig. 2G-2I, display linear $s^2$ dependence over the characteristic Bragg reflections, which agrees well with theoretical prediction. Therefore, these experiments ambiguously evidence the negligible effect of multiple scattering effect to the experimental results.

In our experiment, we carefully check intensity changes of all Bragg reflections in the acquired diffraction pattern and do not observe any signature of photoinduced displacement on the ultrafast time scale. Any relaxation of local strains on longer time scale is also not observed. Therefore, the impact from photoinduced displacement and local strain are excluded.

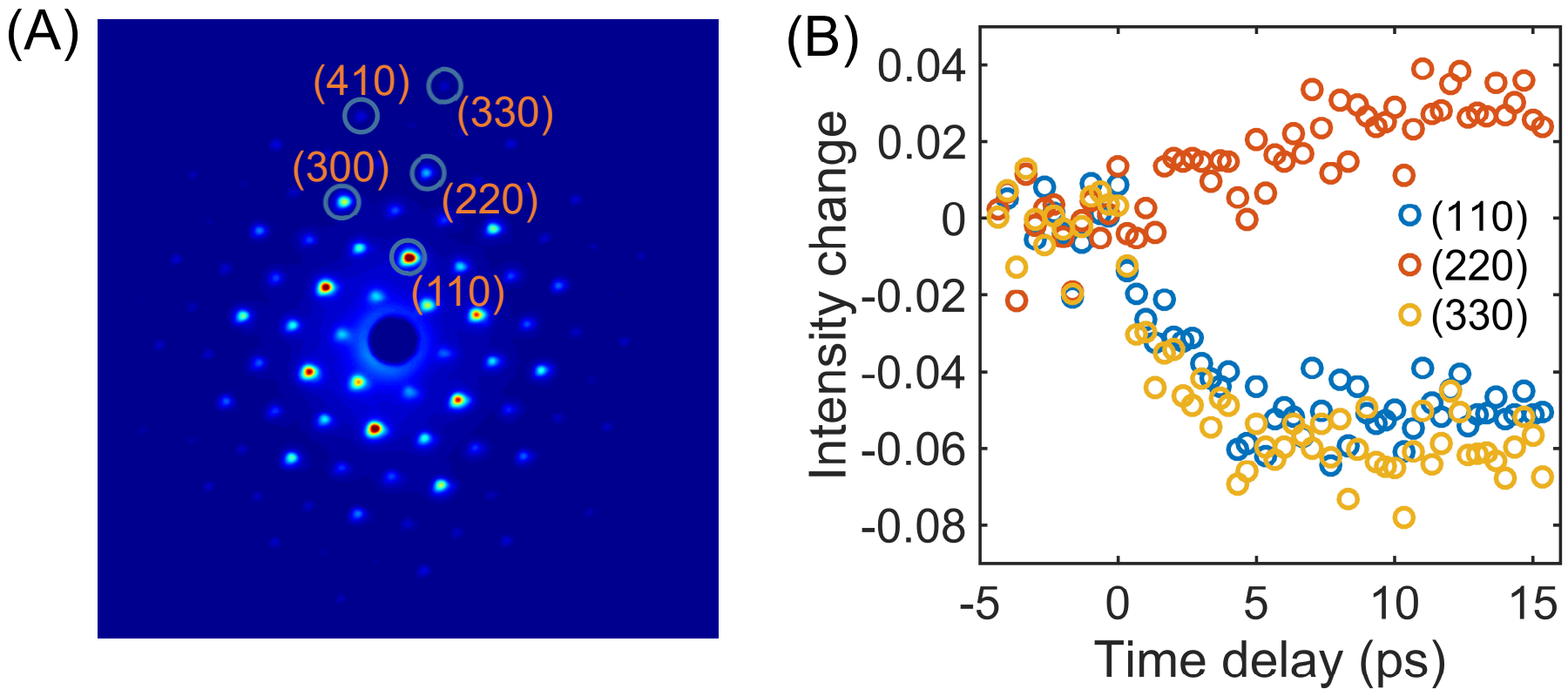

Fig. S2 (A) Typical diffraction pattern obtained in femtosecond electron diffraction experiments with labelled Bragg reflections. (B) Intensity change of the characteristic Bragg peaks as a function of time delay at 11 K. The pump fluence is 3.4 mJ/cm$^2$.

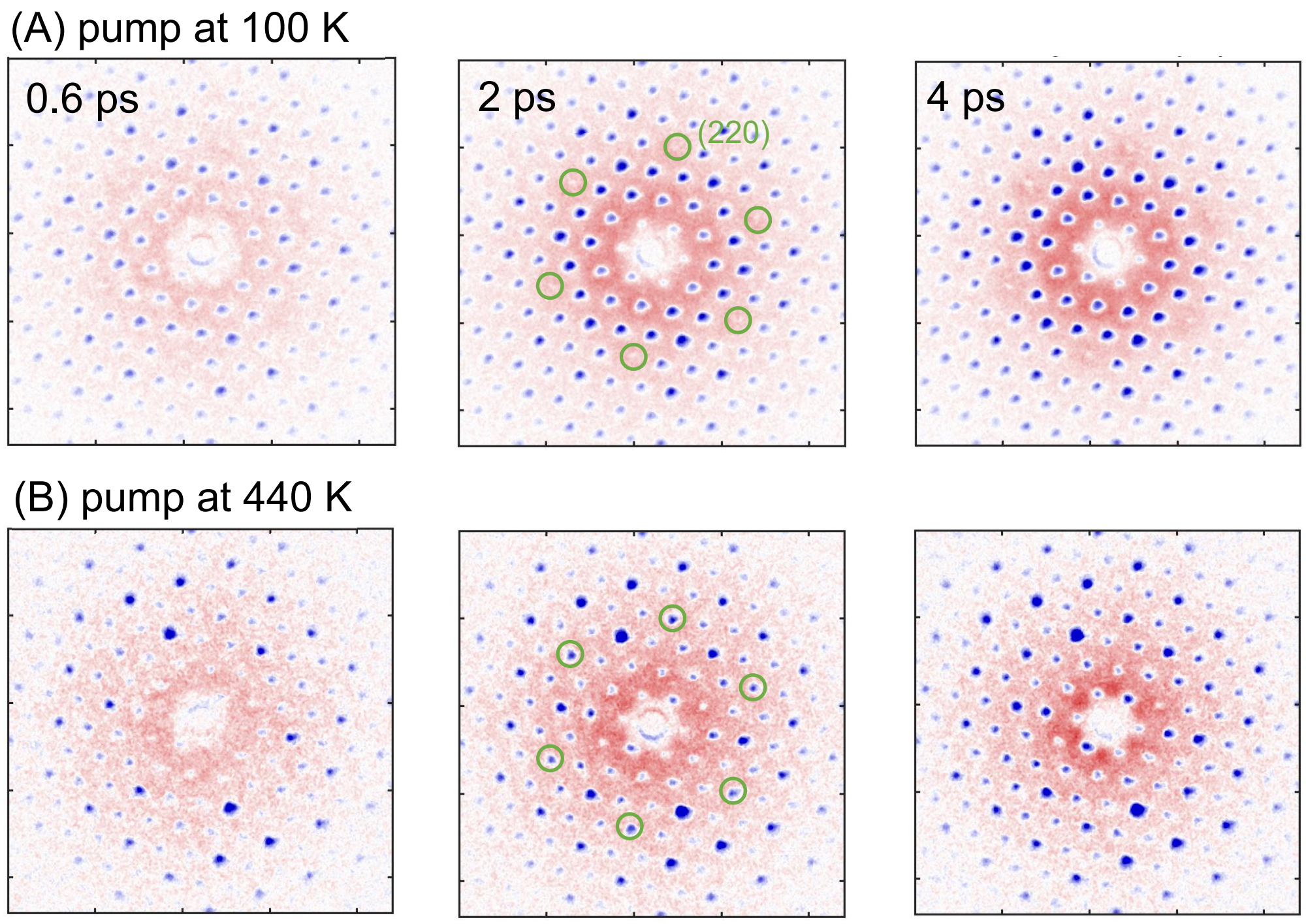

Fig. S3 (A) Differential diffraction pattern at the time delay of 0.6 ps, 2 ps and 4 ps respectively at 100 K. (B) Differential diffraction pattern at the time delay of 0.6 ps, 2 ps and 4 ps respectively at 440 K.

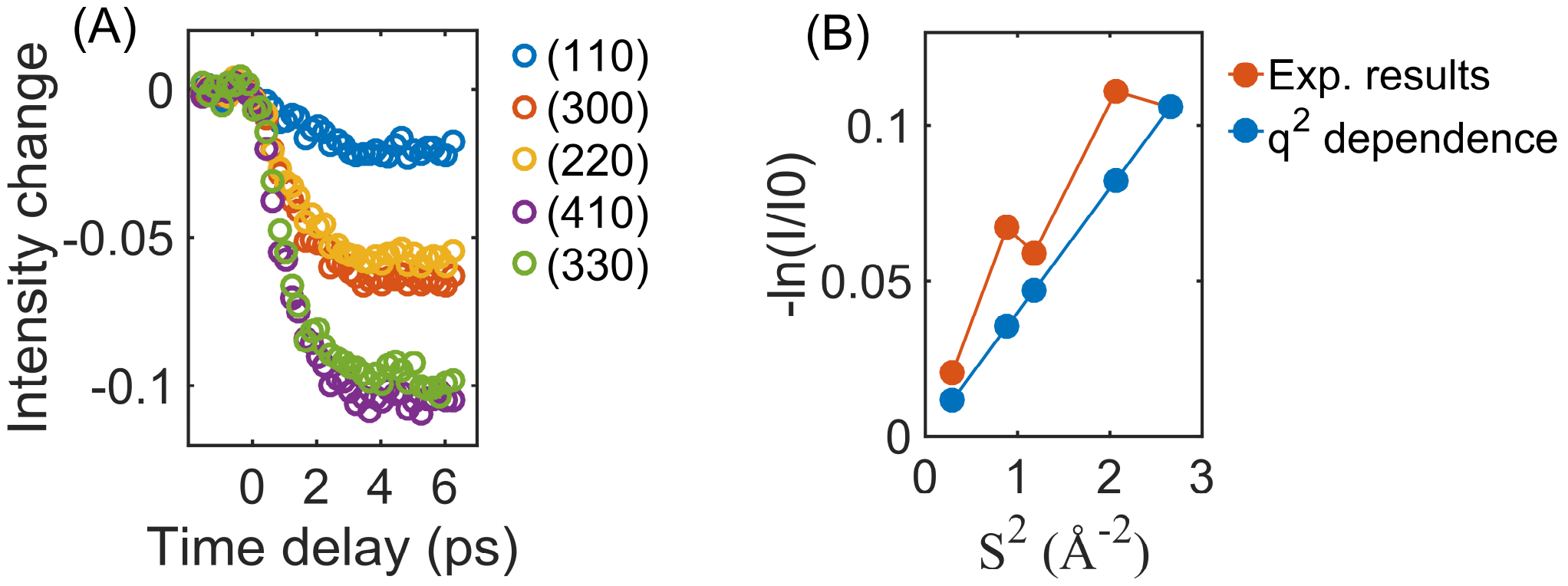

Fig. S4 Ultrafast dynamics of $CuCrSe_2$ measured at base temperature of 200 K. (A) Experimental intensity changes of characteristic Bragg peaks at 200 K; (B) The $s^2$ (i.e., q2) dependences of the experimental results at 200 K and the theoretical prediction results. The pump fluence is 2.6 mJ/$cm^2$.

**Table S1.** The fraction coordinates of $AgCrSe_2$ in the *R3m* phase. In the main text, the (110), (300), (220), (410) and (330) reflections are all characteristic Bragg peaks. We mainly focus on the (110), (220) and (330) peaks because they share the same projection direction for arbitrary atomic displacements and vibrations.

| Atom | *x* | *y* | *z* |
|---|---|---|---|
| $Ag_1$ | 0 | 0 | 0.1529 |
| $Ag_2$ | 0.6667 | 0.3333 | 0.4862 |
| $Ag_3$ | 0.3333 | 0.6667 | 0.8196 |
| $Cr_1$ | 0 | 0 | 0 |
| $Cr_2$ | 0.6667 | 0.3333 | 0.3333 |
| $Cr_3$ | 0.3333 | 0.6667 | 0.6667 |
| $Se_1$ | 0.6667 | 0.3333 | 0.0668 |
| $Se_2$ | 0 | 0 | 0.2694 |
| $Se_3$ | 0.3333 | 0.6667 | 0.4002 |
| $Se_4$ | 0.6667 | 0.3333 | 0.6027 |
| $Se_5$ | 0 | 0 | 0.7335 |
| $Se_6$ | 0.3333 | 0.6667 | 0.9361 |

**Table S2.** The fraction coordinates of $AgCrSe_2$ in the *R-3m* phase. The occupation of the Ag atoms in the table is 0.5.

| Atom | *x* | *y* | *z* |
|---|---|---|---|
| $Ag_1$(0.5) | 0 | 0 | 0.1529 |
| $Ag_2$(0.5) | 0.6667 | 0.3333 | 0.4862 |
| $Ag_3$(0.5) | 0.3333 | 0.6667 | 0.8196 |
| $Ag_4$(0.5) | 0.6667 | 0.3333 | 0.1808 |
| $Ag_5$(0.5) | 0.3333 | 0.6667 | 0.5142 |
| $Ag_6$(0.5) | 0 | 0 | 0.8475 |
| $Cr_1$ | 0 | 0 | 0 |
| $Cr_2$ | 0.6667 | 0.3333 | 0.3333 |
| $Cr_3$ | 0.3333 | 0.6667 | 0.6667 |
| $Se_1$ | 0.6667 | 0.3333 | 0.0668 |
| $Se_2$ | 0 | 0 | 0.2694 |
| $Se_3$ | 0.3333 | 0.6667 | 0.4002 |
| $Se_4$ | 0.6667 | 0.3333 | 0.6027 |
| $Se_5$ | 0 | 0 | 0.7335 |
| $Se_6$ | 0.3333 | 0.6667 | 0.9361 |

## Searching for local correlated structures with the structure factor calculation

The term of local correlated structures suggest static off-symmetry displacements in the subunit cell and these static displacements across unit cells do not form a long-range order. In this case, the averaged symmetry over amount of unit cells is not changed. The organizational forms of these local correlated structures encompass the coexistence of multiple competing phases [38], topological structures [39], quasisymmetry [40,41] and gauge symmetry [42], among others. To simplify the calculation, we perform the structure factor calculation with a single unit cell, rather than a supercell, to model the diffraction intensity change of the overall lattice. Based on the formula (1) in the last section, the structure factor calculation involves the static displacement of Ag atoms and the thermal vibrations of Ag, Cr and Se atoms respectively. Then the standard error (SE) between the experimental and calculated intensity changes over the characteristic Bragg peaks is used to constrain the structural model. The upper limit of SE is set to 0.3 corresponding to a tolerance of ~ ± 30% of the experimental intensity changes, which is within the scope of the intensity resolution of our FED systems. During the search, the amplitude boundary of both the static displacement (≤ 0.37 Å) and the thermal vibration (≤ 0.1 Å) is set to within 10% of the bond length (the Lindemann criteria). To avoid reaching the local minimum instead of the global minimum, the convergence of the search results is always checked. The step size of all parameter changes in the search is dynamically adjusted to ensure the convergence of the search results as well as the calculation efficiency. Based on these sophisticated calculations and analyses, 48 local structures and their corresponding vibration amplitudes are identified as shown in Table S3. Note that, the 48 local atomic configurations are degenerated, all of which giving rise to the same calculated intensity changes over the reflections as uniformly increasing the amplitude of the thermal vibration. In addition, the structure factor calculation with a supercell including multiple unit cells probably give rise to a more precise structural model. This is beyond the scope of the present work due to the limited computation source. As we expand the boundary of the amplitude of the Ag displacements to ~ 0.5 Å and the other parameters remain, 2136 local configurations enable to reproduce the experimental results. The finite number of the local figurations indicates that the local displacements among Ag atoms are correlated and non-ergodic instead of random.

**Table S3.** List of the 48 local structures from simulation. Ag_1x to Ag_3y resemble the displacement in fractional coordinate along x and y axis of the unit cell. The length of the x and y axis is 3.68 Å, then the 0.1 (-0.1) in the table indicates the displacement of 0.37 (-0.37) Å. Ag_amp, Cr_amp and Se_amp resemble the amplitude (Å) of thermal vibration induced by laser excitation.

| Num. | Ag_1x | Ag_1y | Ag_2x | Ag_2y | Ag_3x | Ag3_y | Ag_amp | Cr_amp | Se_amp |
|---|---|---|---|---|---|---|---|---|---|
| 1 | 0.04 | 0.02 | -0.1 | -0.1 | -0.1 | -0.1 | 0.1 | 0.02 | 0.04 |
| 2 | -0.1 | -0.1 | -0.1 | -0.1 | -0.04 | 0.1 | 0.1 | 0.02 | 0.04 |
| 3 | -0.1 | -0.1 | -0.1 | -0.1 | -0.02 | 0.08 | 0.1 | 0.02 | 0.04 |
| 4 | -0.1 | -0.1 | -0.1 | -0.1 | 0 | 0.06 | 0.1 | 0.02 | 0.04 |
| 5 | -0.1 | -0.1 | -0.1 | -0.1 | 0.02 | 0.04 | 0.1 | 0.02 | 0.04 |
| 6 | -0.1 | -0.1 | -0.1 | -0.1 | 004 | 0.02 | 0.1 | 0.02 | 0.04 |
| 7 | -0.1 | -0.1 | -0.1 | -0.1 | 0.06 | 0 | 0.1 | 0.02 | 0.04 |
| 8 | -0.1 | -0.1 | -0.1 | -0.1 | 0.08 | -0.02 | 0.1 | 0.02 | 0.04 |
| 9 | -0.1 | -0.1 | -0.1 | -0.1 | 0.1 | -0.04 | 0.1 | 0.02 | 0.04 |
| 10 | -0.1 | -0.1 | -0.04 | 0.1 | -0.1 | -0.1 | 0.1 | 0.02 | 0.04 |
| 11 | -0.1 | -0.1 | -0.02 | 0.08 | -0.1 | -0.1 | 0.1 | 0.02 | 0.04 |
| 12 | -0.1 | -0.1 | 0 | 0.06 | -0.1 | -0.1 | 0.1 | 0.02 | 0.04 |
| 13 | -0.1 | -0.1 | 0.02 | 0.04 | -0.1 | -0.1 | 0.1 | 0.02 | 0.04 |
| 14 | -0.1 | -0.1 | 0.04 | 0.02 | -0.1 | -0.1 | 0.1 | 0.02 | 0.04 |
| 15 | -0.1 | -0.1 | 0.06 | 0 | -0.1 | -0.1 | 0.1 | 0.02 | 0.04 |
| 16 | -0.1 | -0.1 | 0.08 | -0.02 | -0.1 | -0.1 | 0.1 | 0.02 | 0.04 |
| 17 | -0.1 | -0.1 | 0.1 | -0.04 | -0.1 | -0.1 | 0.1 | 0.02 | 0.04 |
| 18 | -0.1 | 0.04 | 0.1 | 0.1 | 0.1 | 0.1 | 0.1 | 0.02 | 0.04 |
| 19 | -0.08 | 0.02 | 0.1 | 0.1 | 0.1 | 0.1 | 0.1 | 0.02 | 0.04 |
| 20 | -0.06 | 0 | 0.1 | 0.1 | 0.1 | 0.1 | 0.1 | 0.02 | 0.04 |
| 21 | -0.04 | -0.02 | 0.1 | 0.1 | 0.1 | 0.1 | 0.1 | 0.02 | 0.04 |
| 22 | -0.04 | 0.1 | -0.1 | -0.1 | -0.1 | -0.1 | 0.1 | 0.02 | 0.04 |
| 23 | -0.02 | -0.04 | 0.1 | 0.1 | 0.1 | 0.1 | 0.1 | 0.02 | 0.04 |
| 24 | -0.02 | 0.08 | -0.1 | -0.1 | -0.1 | -0.1 | 0.1 | 0.02 | 0.04 |
| 25 | 0 | -0.06 | 0.1 | 0.1 | 0.1 | 0.1 | 0.1 | 0.02 | 0.04 |
| 26 | 0 | 0.06 | -0.1 | -0.1 | -0.1 | -0.1 | 0.1 | 0.02 | 0.04 |
| 27 | 0.02 | -0.08 | 0.1 | 0.1 | 0.1 | 0.1 | 0.1 | 0.02 | 0.04 |
| 28 | 0.02 | 0.04 | -0.1 | -0.1 | -0.1 | -0.1 | 0.1 | 0.02 | 0.04 |
| 29 | 0.04 | -0.1 | 0.1 | 0.1 | 0.1 | 0.1 | 0.1 | 0.02 | 0.04 |
| 30 | 0.06 | 0 | -0.1 | -0.1 | -0.1 | -0.1 | 0.1 | 0.02 | 0.04 |
| 31 | 0.08 | -0.02 | -0.1 | -0.1 | -0.1 | -0.1 | 0.1 | 0.02 | 0.04 |
| 32 | 0.1 | -0.04 | -0.1 | -0.1 | -0.1 | -0.1 | 0.1 | 0.02 | 0.04 |
| 33 | 0.1 | 0.1 | -0.1 | 0.04 | 0.1 | 0.1 | 0.1 | 0.02 | 0.04 |
| 34 | 0.1 | 0.1 | -0.08 | 0.02 | 0.1 | 0.1 | 0.1 | 0.02 | 0.04 |
| 35 | 0.1 | 0.1 | -0.06 | 0 | 0.1 | 0.1 | 0.1 | 0.02 | 0.04 |
| 36 | 0.1 | 0.1 | -0.04 | -0.02 | 0.1 | 0.1 | 0.1 | 0.02 | 0.04 |
| 37 | 0.1 | 0.1 | -0.02 | -0.04 | 0.1 | 0.1 | 0.1 | 0.02 | 0.04 |
| 38 | 0.1 | 0.1 | 0 | -0.06 | 0.1 | 0.1 | 0.1 | 0.02 | 0.04 |
| 39 | 0.1 | 0.1 | 0.02 | -0.08 | 0.1 | 0.1 | 0.1 | 0.02 | 0.04 |
| 40 | 0.1 | 0.1 | 0.04 | -0.1 | 0.1 | 0.1 | 0.1 | 0.02 | 0.04 |
| 41 | 0.1 | 0.1 | 0.1 | 0.1 | -0.1 | 0.04 | 0.1 | 0.02 | 0.04 |
| 42 | 0.1 | 0.1 | 0.1 | 0.1 | -0.08 | 0.02 | 0.1 | 0.02 | 0.04 |
| 43 | 0.1 | 0.1 | 0.1 | 0.1 | -0.06 | 0 | 0.1 | 0.02 | 0.04 |
| 44 | 0.1 | 0.1 | 0.1 | 0.1 | -0.04 | -0.02 | 0.1 | 0.02 | 0.04 |
| 45 | 0.1 | 0.1 | 0.1 | 0.1 | -0.02 | -0.04 | 0.1 | 0.02 | 0.04 |
| 46 | 0.1 | 0.1 | 0.1 | 0.1 | 0 | -0.06 | 0.1 | 0.02 | 0.04 |
| 47 | 0.1 | 0.1 | 0.1 | 0.1 | 0.02 | -0.08 | 0.1 | 0.02 | 0.04 |
| 48 | 0.1 | 0.1 | 0.1 | 0.1 | 0.04 | -0.1 | 0.1 | 0.02 | 0.04 |

**Table S4.** List of the several local structures from simulation. Ag_1x to Ag_3y resemble the displacement in fractional coordinate along x and y axis of the unit cell. The 0.1 (-0.1) in the table indicates the displacement of 0.55 (-0.55) Å. Ag_amp, Cr_amp and Se_amp resemble the amplitude (Å) of thermal vibration induced by laser excitation.

| Num. | Ag_1x | Ag_1y | Ag_2x | Ag_2y | Ag_3x | Ag3_y | Ag_amp | Cr_amp | Se_amp |
|---|---|---|---|---|---|---|---|---|---|
| 1 | -0.1 | -0.06 | -0.1 | -0.02 | 0 | 0.1 | 0.1 | 0.02 | 0.06 |
| 2 | -0.1 | -0.06 | -0.08 | -0.04 | 0 | 0.1 | 0.1 | 0.02 | 0.04 |
| 3 | -0.1 | -0.02 | -0.1 | -0.06 | 0.04 | 0.06 | 0.1 | 0.02 | 0.04 |
| 4 | -0.1 | -0.02 | -0.08 | -0.08 | 0.04 | 0.06 | 0.1 | 0.02 | 0.04 |
| 5 | -0.1 | 0 | 0.04 | 0.1 | 0.08 | 0.06 | 0.1 | 0.08 | 0.04 |

## Density-functional theory (DFT) calculations

The atomic structure of AgCrSe2 was investigated by both machine-learning interatomic potential (MLIP) and density-functional theory (DFT) calculations. First, we construct the auxiliary force constants for $AgCrSe_2$ at different temperatures using the stochastic self-consistent harmonic approximations (SSCHA) method in an iterative way [43,44]. Specifically, 10000 random configurations for a 4×4×4 supercell with 256 atoms following a Bose-Einstein distribution at a temperature T were first constructed from the harmonic force constants and their energies as well as forces were determined by our home-brew MLIP based on neuroevolution potential (NEP) (available on https://github.com/xielin1984/NEP_potentials/tree/main/AgCrSe2). Then the free energy was minimized until a Kong-Liu effective sample size ratio < 0.5 was achieved. During the free energy minimization, the force constants are updated and temperature effect was taken into consideration. New batch of 10000 stochastic configurations was constructed again based on the updated force constants. In such a self-consistent way, a minimization of free energy was achieved with a meaningful factor of 0.5 for gradients.

Once the auxiliary force constants at different temperatures were obtained, we proceeded to a search of polymorphous structures using the anharmonic special displacement methods [45,46]. We first generated supercells with anharnomic special displacements based on the ideal R3m primitive cell and then performed structure relaxations using NEP. It was found that the supercell with anharmonic special displacements would not spontaneously relax back to the ideal high-symmetry R3m phase. Instead, local correlated displacements were formed. As a further confirmation, the relaxed structures were further optimized by DFT calculations using the Vienna ab-initio Software Package (VASP) [47] based on the projected-augmented-wave (PAW) method [48] and PBE exchange-correlation functional [49]. The electron configurations for the Ag, Cr and Se elements were $4d^{10}5s^1$, $3p^63d^54s^1$ and $4s^24p^6$, respectively. An energy cutoff of 500 eV, a Fermi-Dirac smearing of 0.03 eV and a 2×2×2 Monkhorst-Pack mesh were employed during the structure relaxation. Similar to the results obtained by NEP, Ag ions do not relax back to their perfect positions at the center of the Se tetrahedra and would instead form local correlated displacements.

The phonon spectral functions were calculated by unfolding the phonon band structure of the supercell (256 atoms) to the R3m primitive unitcell [50]. Similarly, the electron spectral functions were obtained by unfolding the electronic band structure of the supercell to the R3m primitive unitcell.

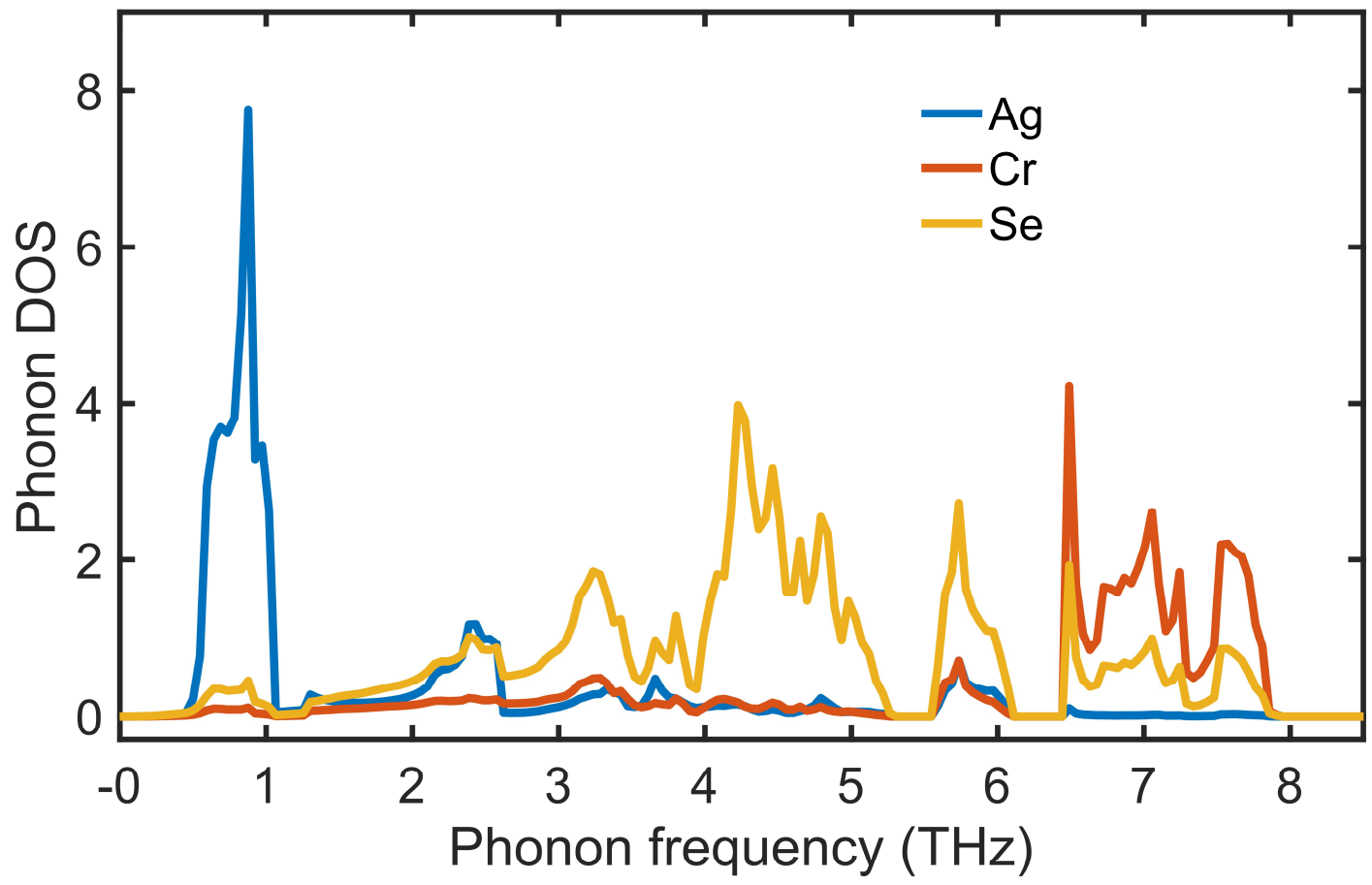

Fig. S5 Phonon density of states (DOS) of $AgCrSe_2$ in the average R3m phase calculated via DFT. Ag atoms dominate the low frequency phonon modes (< 1 THz).

## Macroscopic MLMD calculations and neutron scattering calculations

Machine-learning molecular dynamics (MLMD) were performed using Graphics Processing Units Molecular Dynamics based on neuroevolution potential [51]. The machine-learning potentials utilized in this work can be obtained from publicly available repositories, as reported in Refs. [52]. An orthorhombic supercell was constructed from the conventional hexagonal unit cell of $AgCrSe_2$ with a total number of 17280 atoms. We first performed a 1 ns equilibrium simulation at 300 K in the canonical (NVT) ensemble using the Langevin thermostat with a timestep of 2 fs. Then another 1 ns equilibrium simulation at 10 K in the NVT ensemble was performed, followed by a 500 ps production stage in the microcanonical (NVE) ensemble for collecting atomic trajectories.

The static structure factor of neutron scattering was calculated by the software package LiquidLib (https://linkinghub.elsevier.com/retrieve/pii/S0010465518300778).

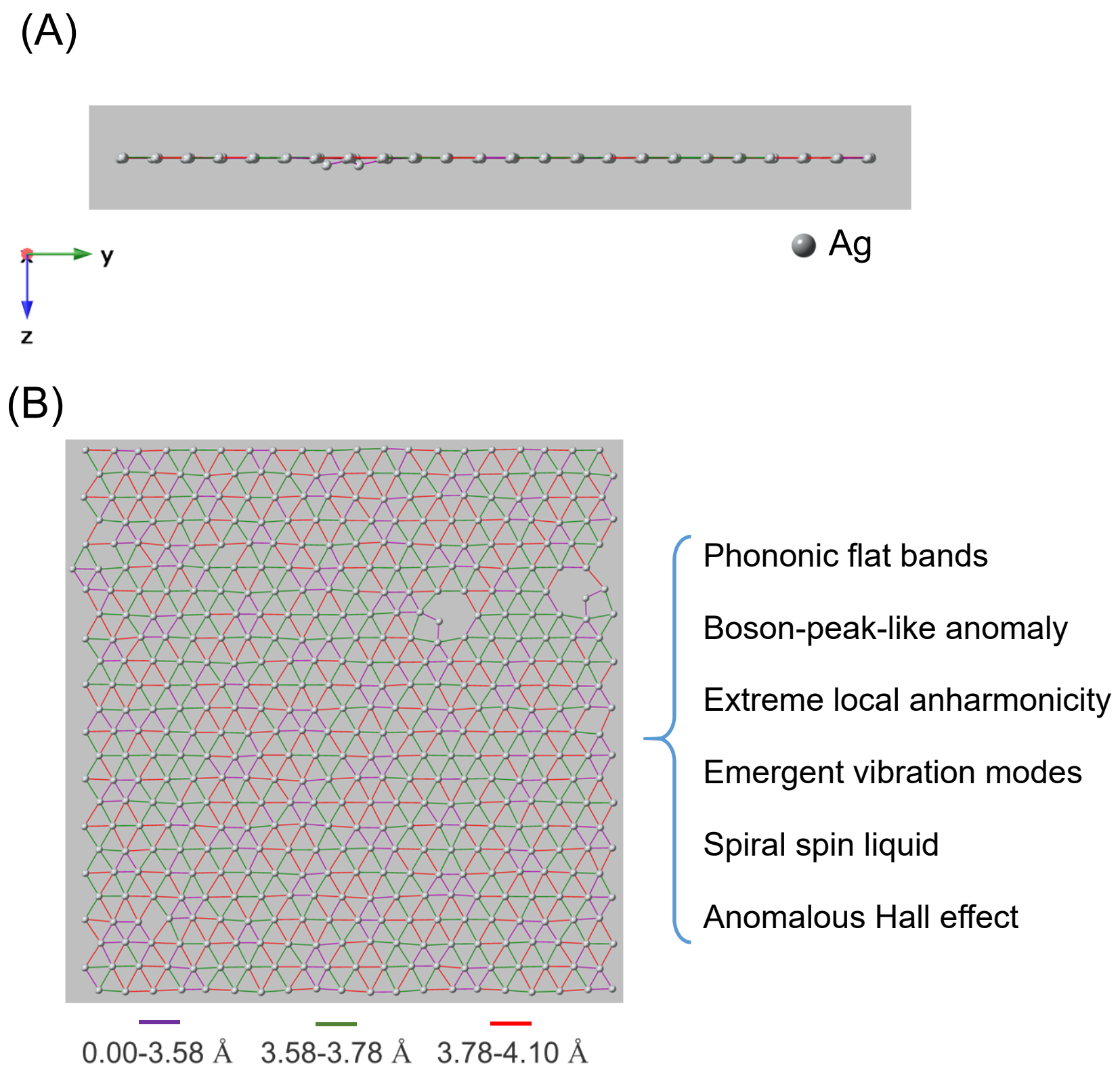

Fig. S6 (A) Cross-sectional view of the single layer Ag distribution along the x direction. The Ag atom displacements are confined to the Ag layers, i.e. the x-y plane. (B) Distribution of Ag-Ag bond lengths. The right panel summarizes the associated anomalous and quantum phenomena arising from many body localization in $AgCrSe_2$.

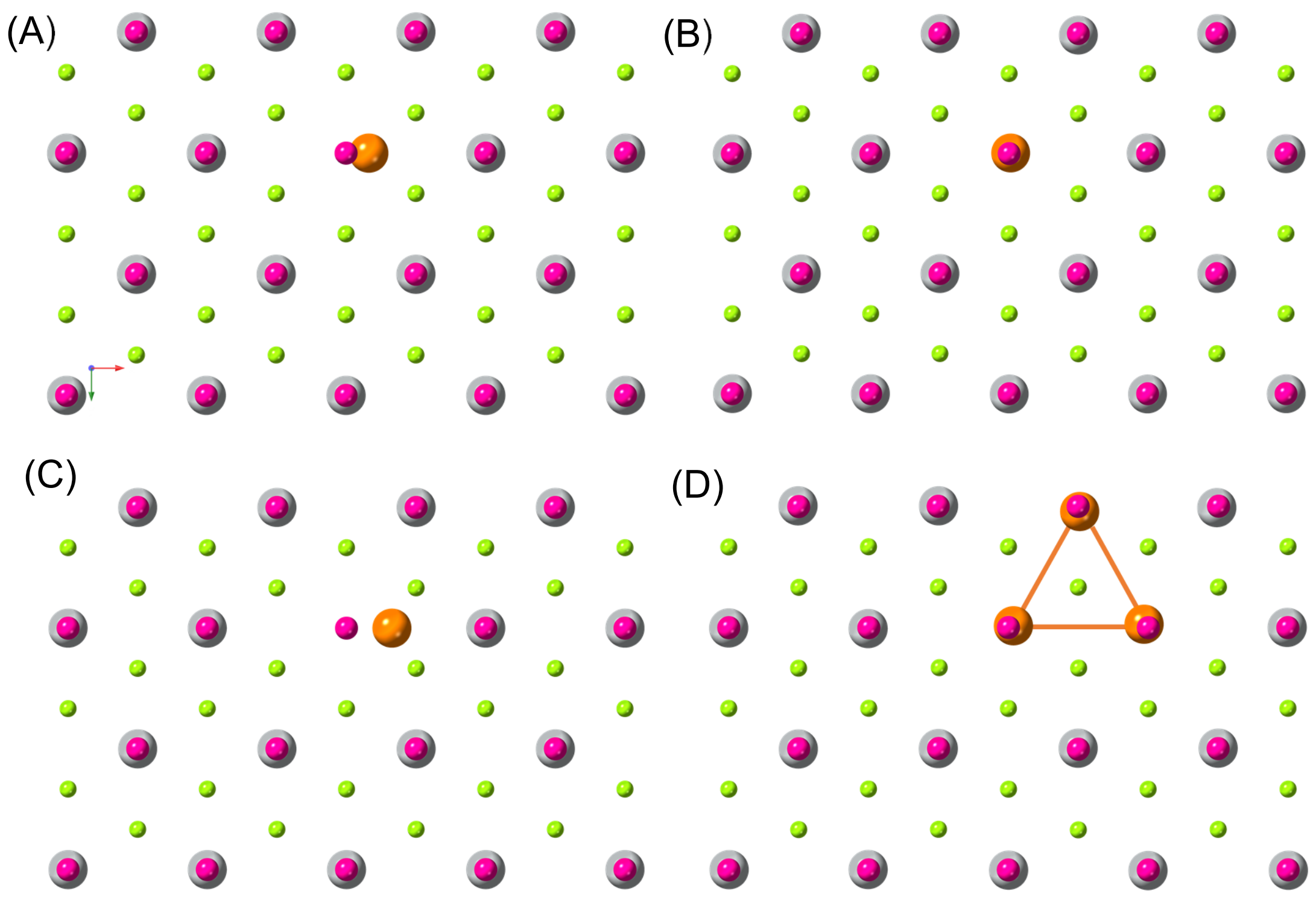

Fig. S7 DFT simulation of the atomic engineering of the DFMBL phase. (A) Horizontal displacement of 0.6 Å of a single Ag atom position with local tweezer. (B) Relaxed structure of the supercell. No significant triangular motif is observed. (C) Horizontal displacement of 1.2 Å of a single Ag atom position with local tweezer. (D) After relaxation, a triangular motif with shorter bond lengths. The displacement of 1.2 Å overcomes an energy barrier to form the triangular motif unattainable with the 0.6 Å displacement.

## Characterizing DFMBL materials with a phononic flat band indicator

Several representative material systems, listed in Table S5, are discussed to elucidate the roles of imaginary phononic flat bands and energy-dependent phononic flat bands in determining DFMBL materials. The DFMBL phase in $AgCrSe_2$ and the absence of the DFMBL phase in $CuCrSe_2$ have illustrated in the main text. Here we firstly use antiferromagnetic $NiPS_3$ to investigate whether high-energy phononic flat bands induce the DFMBL phase. Fig. S8A and Fig. S8B displays the diffraction pattern and time-resolved intensity changes of characteristic Bragg reflections ((060), (0120) and (0180)). In Fig. S8C, the intensity changes of these peaks scale with $q^2$, i.e., in a 1:4:9 ratio, confirming an idealized average crystal structure. Such structural responses remain consistent as the temperature varies from 15 K to ambient conditions. Therefore, no signature of DFMBL phase displays experimentally in $NiPS_3$. The phonon spectra function in Fig. S9A, shows significant phononic flat bands around 7.5 THz, 11 THz and 12.5 THz. In contrast to the low-energy phononic flat bands in $AgCrSe_2$, these high-energy phononic flat bands in $NiPS_3$ do not support a DFMBL phase. Using the same structure relaxation approach as for $AgCrSe_2$, the relaxed $NiPS_3$ structure in Fig. S9C adopts the same structure as the idealized average structure (Fig. S9B), thereby theoretically excluding a DFMBL phase.

As reported recently in our study [53], robust sub-angstrom localization appears to exist in sub-unit cell in $BaTiS_3$. The phonon spectra function of this material exhibits significant imaginary phononic flat bands associated with the these local displacements (see Fig. 1B in [53]). Therefore, imaginary phononic flat bands suggest a good indicator of a DFMBL phase.

Two additional examples, InTe with imaginary phononic flat bands and $CuInP_2S_6$ with real phononic flat bands, are presented in Table S5, suggesting a DFMBL phase in these materials and warranting further experimental verification. Consequently, based on these criteria, a catalogue of naturally occuring DFMBL materials is anticipated through database screening of phonon properties for inorganic crystals (see a typical database at https://chemrxiv.org/doi/full/10.26434/chemrxiv.15004632/v1).

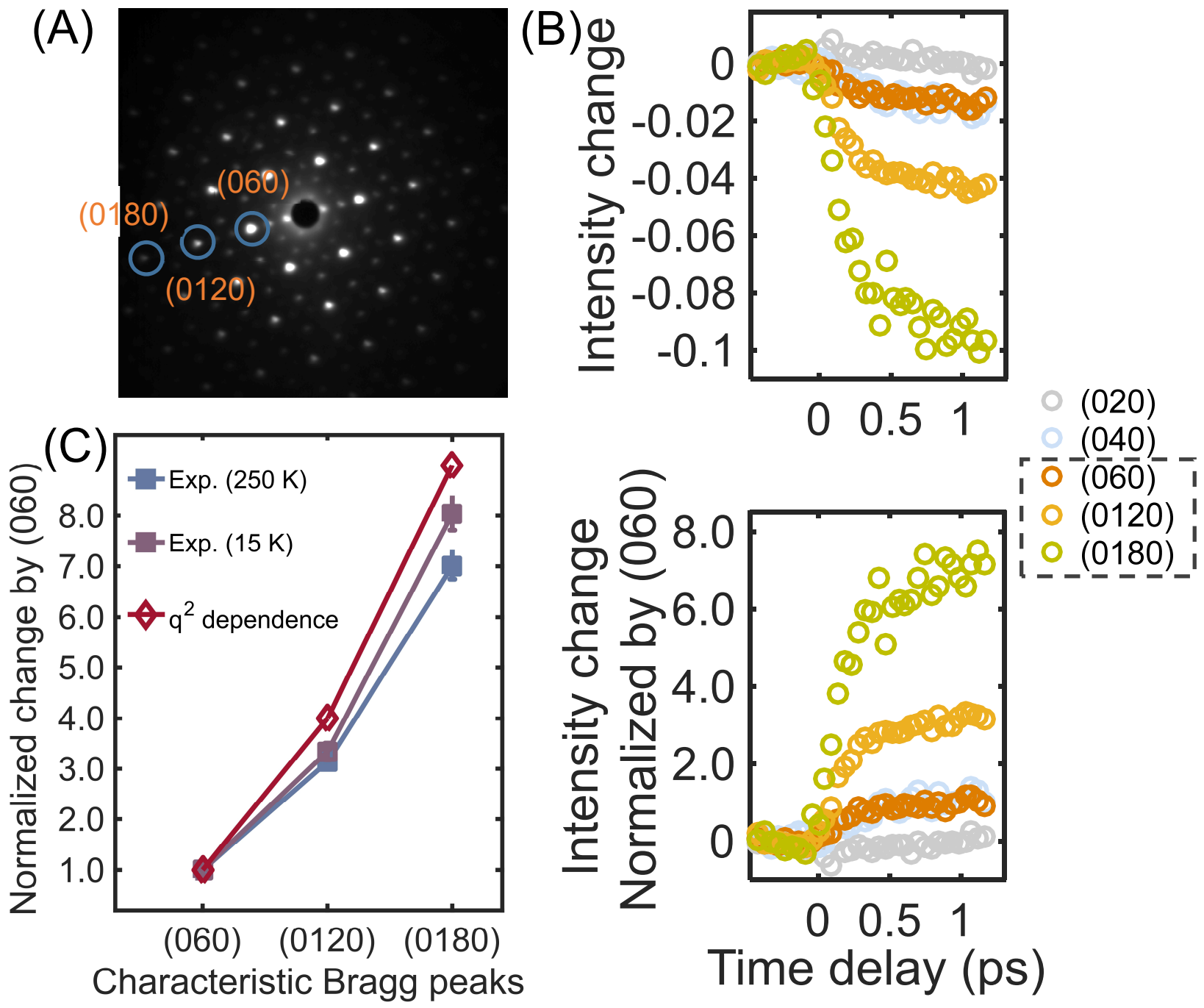

Fig. S8 (A) Diffraction pattern obtained in femtosecond electron diffraction experiments with labelled characteristic Bragg reflections. (B) Intensity change of the Bragg peaks as a function of time delay at 15 K. (C) Summary of the intensity changes normalized by the change of (060). The pink squares represent the theoretical $q^2$ dependence (1:4:9).

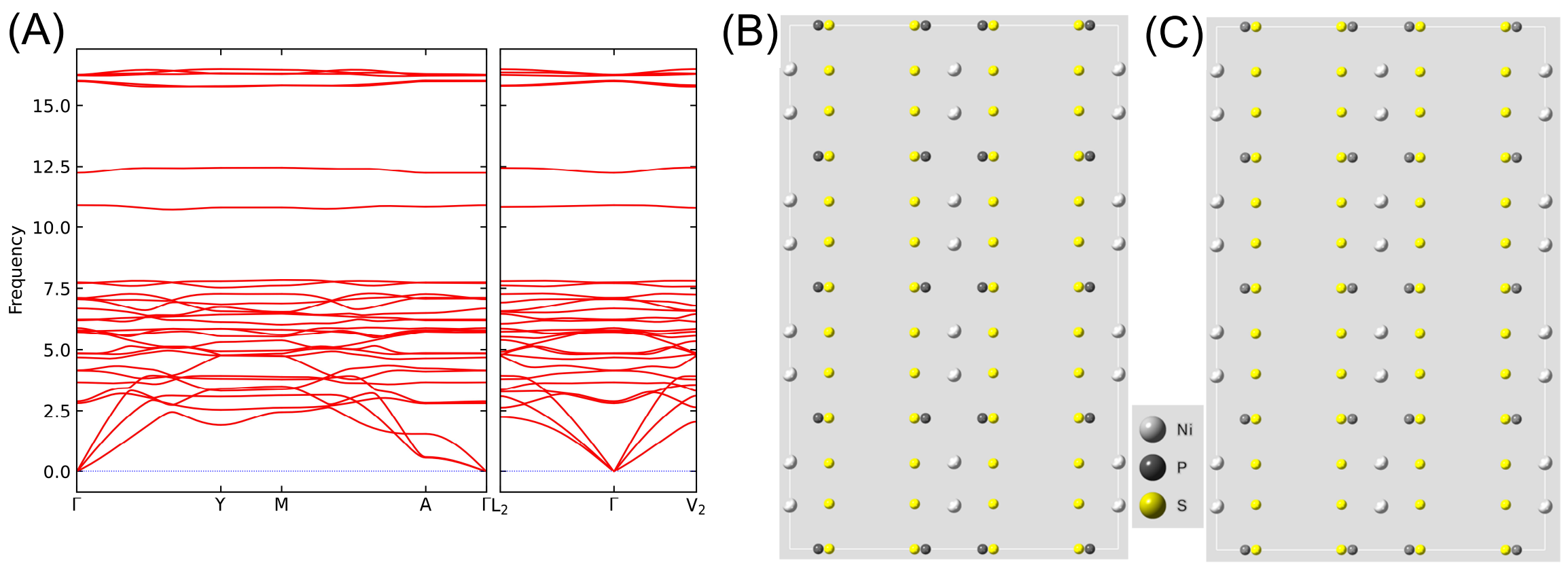

Fig. S9 (A) Phonon spectral function of $NiPS_3$ along the high symmetry k-path. (B) Idealized average structure of $NiPS_3$. (C) Relaxed structure at 600 K using neuroevolution potential.

**Table S5.** Representative materials to characterize the DFMBL phase. The first four have been identified and the last two warrant further experimental verification.

| | | | |
|---|---|---|---|
| $AgCrSe_2$ | Low-energy flat bands | DFMBL phase | This work |
| $CuCrSe_2$ | No flat bands | No DFMBL phase | This work |
| $NiPS_3$ | High-energy flat bands | No DFMBL phase | This work |
| $BaTiS_3$ | Imaginary flat bands | DFMBL phase | Our work Ref. [53] |
| InTe | Imaginary flat bands | Candidate of DFMBL phase | Ref. [54] |
| $CuInP_2S_6$ | Low-energy flat bands | Candidate of DFMBL phase | Ref. [55] |

**Functional materials with characteristic Bragg peaks**

As far as we know, many quantum materials and functional materials exhibit characteristic Bragg reflections, therefore, the possible local structures in these material systems could be identified by the Bragg scattering phase breaking regime. The table below display some representative materials and relevant research is currently underway. Our recent work identifies the local disorder in SnSe [35], while the intrinsic mechanism in revealing the local structures by ultrafast dynamics of Bragg reflections is not clear, which is first proposed in the present work.

**Table S6.** Representative functional and quantum materials with characteristic Bragg peaks

| | | |
|---|---|---|
| SnSe | thermoelectricity | (020) (040) (060) and high order peaks |
| $AgCrSe_2$ | Superionic conductor | (110) (220) (330) and high order peaks |
| $TiSe_2$ | Charge density wave | (030) (060) (090) and high order peaks |
| 2H-$MoS_2$ | 2D semiconductor | (030) (060) (090) and high order peaks |
| 2M-$WS_2$ | Strange metal | (020) (040) (060) and high order peaks |
| FeSe | superconductivity | (020) (040) (060) and high order peaks |
| hBN | 2D dielectric | (030) (060) (090) and high order peaks |
| Ni | magnetism | (200) (400) (600) and high order peaks |
| Black phosphorus | 2D semiconductor | (200) (400) (600) and high order peaks |
| $Co_3Sn_2S_2$ | Kagome lattice | (30) (600) (900) and high order peaks |